\documentclass[letterpaper,twocolumn,10pt]{article}
\usepackage{usenix2019_v3}

\usepackage{tikz}
\usepackage{amsmath}

\usepackage{xspace}
\usepackage{listings}
\usepackage{multirow}
\usepackage{array}
\usepackage[ruled,linesnumbered]{algorithm2e}
\usepackage{algpseudocode}
\usepackage[nodisplayskipstretch]{setspace}
\usepackage{caption}
\usepackage[skip=0pt]{subcaption}
\usepackage{amsfonts}
\usepackage[english]{babel}
\usepackage{url}
\usepackage[export]{adjustbox}
\usepackage{wasysym}

\usepackage{amsthm}
\theoremstyle{definition}

\newcommand{\system}{\textsc{CAD}\xspace}
\newcommand{\testbed}{MCity\xspace}

\newcommand{\tool}[1]{#1}

\newcommand{\BULLET}{\vspace{+.00in} \noindent {\footnotesize$\bullet$} \hspace{+.00in}}

\newcommand{\ie}{\textit{i.e.,}\xspace}
\newcommand{\eg}{\textit{e.g.,}\xspace}
\newcommand{\etal}{\textit{et al.}\xspace}

\newcommand{\mysection}[1]{\vspace{-.17in}\section{#1}\vspace{-.08in}}
\newcommand{\mysubsection}[1]{\vspace{-.17in}\subsection{#1}\vspace{-.05in}}
\newcommand{\mysubsubsection}[1]{\vspace{-.13in}\subsubsection{#1}\vspace{-.05in}}

\newif\ifshowcomment
\showcommentfalse

\newif\ifshowhighlight
\showhighlightfalse

\ifshowcomment
    \newcommand{\qz}[1]{[{\color{cyan}Qingzhao: #1}]}
    \newcommand{\jiachen}[1]{{\color{blue}{JC: #1}}}
    \newcommand{\alfred}[1]{{\color{blue}{Alfred: #1}}}
\else
    \newcommand{\qz}[1]{}
    \newcommand{\jiachen}[1]{}
    \newcommand{\alfred}[1]{}
\fi

\ifshowhighlight
    \newcommand{\m}[1]{\textcolor{red}{#1}}
\else
    \newcommand{\m}[1]{#1}
\fi

\makeatletter
\newcommand\notsotiny{\@setfontsize\notsotiny\@vipt\@viipt}
\makeatother

\newcommand{\ignore}[1]{}

\begin{document}

\date{}

\title{\Large \bf On Data Fabrication in Collaborative Vehicular Perception:\\Attacks and Countermeasures}

\author{
{\rm Qingzhao Zhang$^1$, Shuowei Jin$^1$, Ruiyang Zhu$^1$, Jiachen Sun$^1$, Xumiao Zhang$^1$,}\\{\rm Qi Alfred Chen$^2$, Z. Morley Mao$^{1,3}$}\\
$^1$University of Michigan, $^2$University of California, Irvine, $^3$Google
}

\maketitle
\pagestyle{empty}

\begin{abstract}
Collaborative perception, which greatly enhances the sensing capability of connected and autonomous vehicles (CAVs) by incorporating data from external resources, also brings forth potential security risks. CAVs' driving decisions rely on remote untrusted data, making them susceptible to attacks carried out by malicious participants in the collaborative perception system. However, security analysis and countermeasures for such threats are absent.
To understand the impact of the vulnerability, we break the ground by proposing various real-time data fabrication attacks in which the attacker delivers crafted malicious data to victims in order to perturb their perception results, leading to hard brakes or increased collision risks. Our attacks demonstrate a high success rate of over 86\% on high-fidelity simulated scenarios and are realizable in real-world experiments.
To mitigate the vulnerability, we present a systematic anomaly detection approach that enables benign vehicles to jointly reveal malicious fabrication. It detects 91.5\% of attacks with a false positive rate of 3\% in simulated scenarios and significantly mitigates attack impacts in real-world scenarios.
\end{abstract}



\mysection{Introduction}

The perception system of connected and autonomous vehicles (CAVs) is safety-critical as its performance directly affects driving decisions~\cite{apollo, autoware}. However, CAV's perception is confronted with the basic limitation that onboard sensors have limited sensing capabilities. For instance, LiDAR, the commonly adopted 3D sensor, cannot see through occlusions and may render low resolutions for far-away objects, leading to imperfect detection performance. Many recent efforts have proposed LiDAR-based collaborative perception algorithms~\cite{chen2019f, wang2020v2vnet, zhang2021emp, xu2022opv2v}, where different nearby vehicles exchange perception information (e.g., raw sensor data or feature maps processed by neural networks) and perform object detection algorithms on the fused data. 
\m{
In terms of the accuracy of object detection, the approach significantly outperforms the traditional CAV collaboration~\cite{gunther2016realizing,schiegg2020analytical,godoy2021grid} sharing simple GPS messages or object locations, as illustrated in related studies~\cite{xu2022opv2v,wang2020v2vnet}.
}
CAV industry~\cite{intel,qualcomm,huawei,yu2022dair,ford,bosch,infineon} also proposes solutions of collaborative perception and launch road testing across the globe.

Although collaborative perception is evolving quickly towards maturity, it introduces a severe vulnerability to vehicle safety because the safety-critical perception algorithms now rely on sensor data or feature maps from remote untrusted vehicles. With the control of a remote vehicle via physical access to either software or hardware, an attacker can fabricate the data to share, aiming to inject fake object detection results into the view of victim vehicles and even mislead them to trigger accidents. 
\m{
However, the impact of such a severe data integrity threat has not been comprehensively evaluated.
Existing studies of CAV security~\cite{pham2021survey,zhang2022design} either focus on other scopes (\eg physical sensor security~\cite{cao2019adversarial,sun2020towards}, network protocols~\cite{baee2019broadcast,yoshizawa2019survey}) or assume a different threat model (\eg single-vehicle perception~\cite{sun2020towards,liu2021seeing}, object-sharing collaboration~\cite{zhao2021detection,ambrosin2019design}), thus existing mitigation methods are not effectively designed for the new threat.
}

To bridge the gap, we propose a series of stealthy, targeted, and realistic attacks exploiting LiDAR-based collaborative perception in this study. Our proposed attacks can spoof or remove objects at specified locations in the victim's perception results, making all mainstream types of collaborative perception schemes vulnerable.
For early-fusion systems which directly merge LiDAR point clouds, we propose black-box ray casting to reconstruct malicious but natural raw point clouds. We design offline adversarial object generation and run-time occlusion-aware point sampling to further optimize the distribution of modified points.
For intermediate-fusion systems which merge feature maps as intermediate results of object detection models, we design a white-box adversarial attack to perturb the feature maps. For optimal efficiency, the adversarial attack initializes the perturbation vector via a black-box method and runs one-step backward propagation in each LiDAR cycle (\eg 100 ms).
More importantly, we propose zero-delay attack scheduling to make attacks realizable in the real world. To be specific, in order to attack the perception of frame $i$, attackers prepare a fabrication plan based on the knowledge of frame $i-1$ before the next frame comes. In this way, attackers earn one LiDAR cycle time to complete attack generation without introducing a noticeable delay in the fabricated data.

We evaluate the attack effectiveness on 211 traffic scenarios in a simulated dataset \tool{Adv-OPV2V} and a real-world dataset \tool{Adv-\testbed} (including 8 scenarios collected from a real-vehicle testbed \testbed~\cite{mcity}).
On the simulated dataset, all attacks have a success rate of more than 86\% regardless of fusion methods and model configurations. In our real-world experiments, we deploy three vehicles equipped with LiDAR/GPS sensors and the latest Baidu Apollo autonomous driving software~\cite{apollo}. Our attacks can be launched in real-time and trigger safety hazards such as collisions and emergent hard brakes.
We also provide a comprehensive analysis of how the attack effectiveness is affected by various factors including attack methods, fusion schemes, and scenarios. Our findings will guide system designers to build robust collaborative perception schemes.

To mitigate the demonstrated attacks, we propose Collaborative Anomaly Detection (\system), a system that detects data fabrication attacks by revealing geometry inconsistencies of the shared data from different vehicles.
To achieve this, \system requires each vehicle to generate and share an occupancy map, which is a 2D map labeling the 2D space into three classes, free, occupied, and unknown. On receiving occupancy maps from others, the vehicle validates the consistency of the maps, \ie there is no region classified as occupied and free at the same time. 
Then the vehicle carries out the second check by merging the occupancy maps into one and checking perception results against it. For instance, free regions should not overlap with detected bounding boxes; each on-road moving occupied region should have one bounding box overlap with it.
In this way, abnormal detection results caused by either fabricated data or perception faults are revealed \m{if the attacked region is observed by at least one benign CAV}.
\system detects 91.5\% attacks with a false positive rate <3\% on datasets \tool{Adv-OPV2V} and \tool{Adv-\testbed}.

As the first comprehensive security analysis of collaborative perception, we will open-source all the above attack/defense practices as a benchmark tool to facilitate future research. Our contributions can be summarized as three-fold:

\BULLET We compile the benchmark datasets \tool{Adv-OPV2V} and \tool{Adv-\testbed} for evaluating the security of collaborative perception. Especially, \tool{Adv-\testbed} is the \textit{first} multi-vehicle collaboration dataset collected on real vehicles and real roads.

\BULLET We propose multiple data fabrication attacks, where one attacker, as a collaborative perception participant, can successfully spoof or remove objects at specified locations. We conduct an extensive study on the impact of such attacks.

\BULLET We develop \system, \m{a defense system of collaborative perception for detecting our proposed data fabrication attacks}. \system reveals abnormal perception results through the sharing of fine-grained occupancy maps.

\mysection{Background and Related Work}
\label{sec:background}

\begin{table}[t]
  \tiny
  \centering
  \caption{\m{Existing attacks on LiDAR collaborative perception.}}
  \vspace{-0.04in}
  \label{tab:related_attacks}
  \setlength{\tabcolsep}{1pt}
  {\centering
  \begin{tabular}{| c | c | c | c | c | c | c |}
    \noalign{\global\arrayrulewidth1pt}\hline\noalign{\global\arrayrulewidth0.4pt}
    Method & Targeted system & Requirements & Spoof & Remove & Targeted & Real-time \\
    \noalign{\global\arrayrulewidth1pt}\hline\noalign{\global\arrayrulewidth0.4pt}
    LiDAR spoofing~\cite{cao2019adversarial,sun2020towards,canilho2016multi,cao2021invisible} & Single-vehicle & Laser emitters & $\CIRCLE$ & $\CIRCLE$ & $\CIRCLE$ & $\CIRCLE$ \\
    Physical objects~\cite{tu2020physically,zhu2021can} & Single-vehicle & Physical access to the target & $\Circle$ & $\CIRCLE$ & $\CIRCLE$ & $\CIRCLE$ \\
    False object messages~\cite{hadded2020security} &  Late-fusion & None & $\CIRCLE$ & $\CIRCLE$ & $\CIRCLE$ & $\CIRCLE$ \\
    Multi-agent adversarial~\cite{tu2021adversarial} & Int.-fusion & Local computation & $\CIRCLE$ & $\CIRCLE$ & $\Circle$ & $\Circle$ \\
    Ours & \textbf{Early/Int.-fusion} & Local computation & $\CIRCLE$ & $\CIRCLE$ & $\CIRCLE$ & $\CIRCLE$ \\
    \noalign{\global\arrayrulewidth1pt}\hline\noalign{\global\arrayrulewidth0.4pt}
  \end{tabular}
  }
  {\tiny \textbf{Notes:} \emph{Int.} - intermediate-fusion.}
\end{table}

\begin{table}[t]
  \tiny
  \centering
  \caption{\m{Effectiveness of defenses on our attacks.}}
  \vspace{-0.04in}
  \label{tab:related_defenses}
  \setlength{\tabcolsep}{1pt}
  \begin{tabular}{| c | c | c | c | c | c | c | c |}
    \noalign{\global\arrayrulewidth1pt}\hline\noalign{\global\arrayrulewidth0.4pt}
    \multirow{2}{*}{Method} & \multirow{2}{*}{Threat model} & \multicolumn{2}{c|}{Attack types} & \multicolumn{3}{c|}{Leveraged information} & \multirow{2}{*}{Overall} \\
    \cline{3-7}
    & & Spoof & Remove & Spatial & Temporal & Multi-vehicle & \\
    \noalign{\global\arrayrulewidth1pt}\hline\noalign{\global\arrayrulewidth0.4pt}
    Network integrity~\cite{IEEE_1609_2} & Cyber attacks & $\Circle$ & $\Circle$ & $\Circle$ & $\Circle$ & $\Circle$ & $\Circle$ \\
    Trusted execution~\cite{hu2020cvshield} & Malicious software & $\CIRCLE$ & $\CIRCLE$ & $\Circle$ & $\Circle$ & $\Circle$ & $\LEFTcircle$ \\
    Occlusion-aware~\cite{sun2020towards} & Physical LiDAR spoofing & $\CIRCLE$ & $\Circle$ & $\CIRCLE$ & $\Circle$ & $\Circle$ & $\LEFTcircle$ \\
    Temporal check~\cite{liu2021seeing} & Physical LiDAR spoofing & $\CIRCLE$ & $\CIRCLE$ & $\LEFTcircle$ & $\CIRCLE$ & $\Circle$ & $\Circle$ \\
    Multi-sensor check~\cite{liu2021seeing} & Physical LiDAR spoofing & $\CIRCLE$ & $\CIRCLE$ & $\Circle$ & $\Circle$ & $\Circle$ & $\LEFTcircle$ \\
    Spatial conflicts w/ ego~~\cite{ambrosin2019design} & Fake message in late-fusion & $\CIRCLE$ & $\CIRCLE$ & $\CIRCLE$ & $\Circle$ & $\LEFTcircle$ & $\LEFTcircle$ \\
    Ours (\system) & Our attacks & $\CIRCLE$ & $\CIRCLE$ & $\CIRCLE$ & $\Circle$ & $\CIRCLE$ & $\CIRCLE$ \\
    \noalign{\global\arrayrulewidth1pt}\hline\noalign{\global\arrayrulewidth0.4pt}
  \end{tabular}
\end{table}

\noindent
\textbf{Connected and autonomous vehicles (CAVs)} are transforming the transportation systems by enabling automatic and intelligent vehicle driving control. CAVs are complicated cyber-physical systems equipped with sensors such as LiDAR, camera, and radar to perceive the surroundings, and software to make appropriate driving decisions. By the end of 2022, numerous companies including Waymo, Honda, Baidu, and Tesla~\cite{waymo,honda,apollo,autopilot} have developed models of CAVs.

\begin{figure*}[!htb]
\minipage{0.36\textwidth}
    \includegraphics[width=\textwidth]{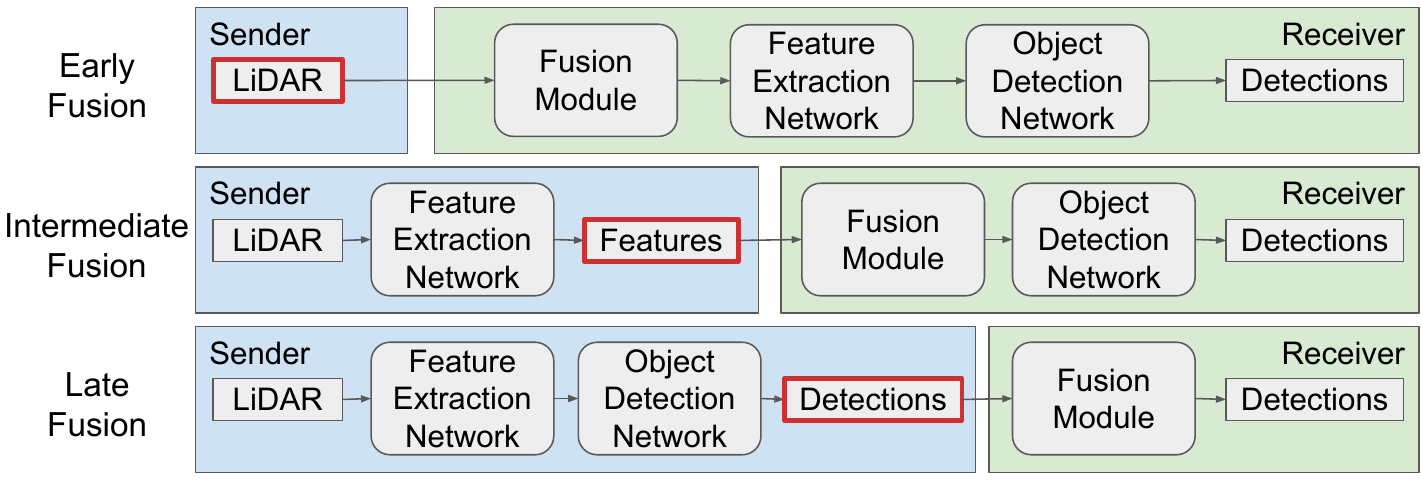}
    \vspace{-0.08in}
    \caption{Taxonomy of cooperative perception systems.}
    \vspace{-0.10in}
    \label{fig:perception}
\endminipage\hfill
\minipage{0.26\textwidth}
    \includegraphics[width=\textwidth]{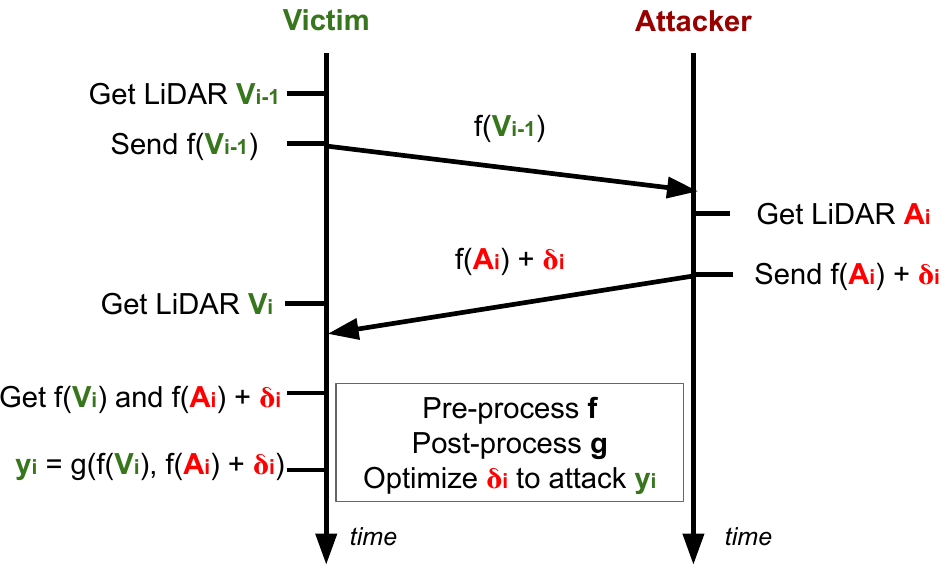}
    \vspace{-0.25in}
    \caption{Temporal order of message exchanges.}
    \vspace{-0.10in}
    \label{fig:temporal_constraint}
\endminipage\hfill
\minipage{0.36\textwidth}
    \includegraphics[width=\textwidth]{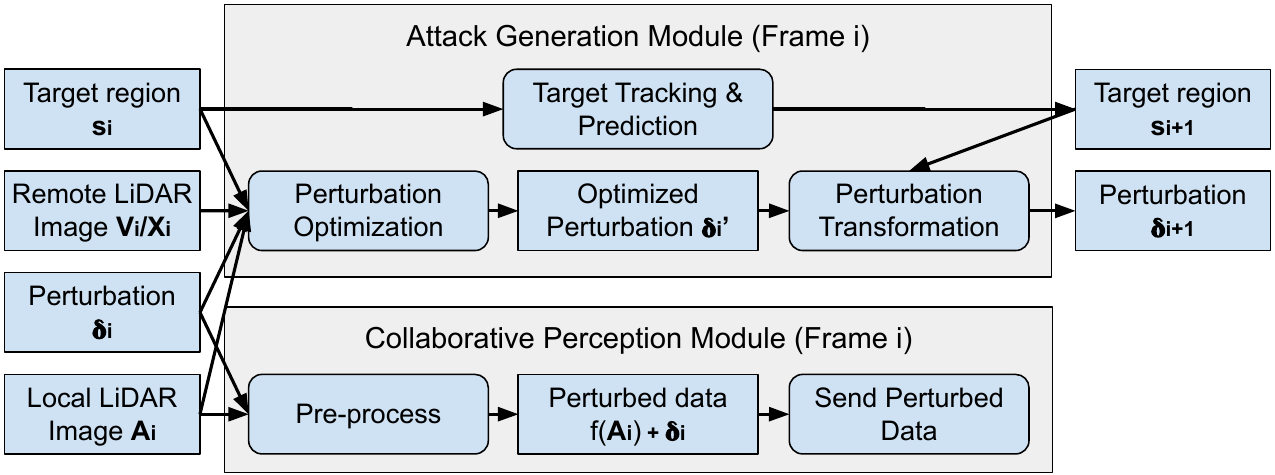}
    \vspace{-0.10in}
    \caption{Framework of real-time targeted attacks on collaborative perception.}
    \vspace{-0.10in}
    \label{fig:attack_overview}
\endminipage
\end{figure*}

\noindent
\textbf{Collaborative perception} has been proposed to enhance CAV perception~\cite{lang2019pointpillars,shi2019pointrcnn,ku2018joint,alaba2022survey,li2020lidar} by sharing raw or processed sensor data among infrastructure or vehicles. Mainstream solutions focus on LiDAR sensors because of the rich 3D geometry features brought by LiDAR images.
Collaborative perception has three major types according to the sharing data, as shown in Figure~\ref{fig:perception}. CAVs in early-fusion sharing schemes~\cite{kumar2012carspeak, chen2019cooper, zhang2021emp, qiu2021autocast, chen2022cooperative, zhang2023robust} directly exchange raw sensor data, whose format is usually universal and can be naively concatenated, at a cost of data transmission bandwidth; intermediate-fusion schemes~\cite{chen2019f, cui2022coopernaut, yuan2022keypoints, wang2020v2vnet, xu2022v2x} ask CAVs to transmit feature maps, the intermediate product of perception algorithms, offering a good tradeoff between network efficiency and perception accuracy; in late-fusion schemes~\cite{liu2019fusioneye, shi2022vips, song2022efficient} lightweight perception results such as object bounding boxes are shared.

Collaborative perception is advancing quickly towards real-world deployment. 3GPP standardized for Cellular Vehicle-to-Everything (C-V2X) techniques in 2017~\cite{3gpp}, indicating the maturity of roadside communication. Since then, major technology companies such as Huawei, Intel, Bosch, Infineon, and Qualcomm~\cite{huawei,intel,bosch,infineon,qualcomm} have strived to build various C-V2X solutions. Road trials have been launched across the globe in countries like Germany, France, the United States, and Japan. 
Ford~\cite{ford} and Baidu Apollo~\cite{yu2022dair} built real-world collaborative perception datasets.

\noindent
\textbf{Attacks on CAV perception}. \m{
Several attacks can harm LiDAR perception systems as listed in Table~\ref{tab:related_attacks}. First, LiDARs on CAVs are vulnerable to physical attacks, such as GPS spoofing~\cite{li2021fooling,shen2020drift}, LiDAR spoofing~\cite{jin2022pla,hallyburton2022security,li2021fooling,cao2021invisible}, and physical realizable adversarial objects~\cite{tu2020physically,zhu2021can,zhang2022adversarial}. These attacks are against one single autonomous vehicle.
Late-fusion collaborative perception shares object locations~\cite{gunther2016realizing,schiegg2020analytical,godoy2021grid,hadded2020security} thus the attacker can trivially modify these locations, which is the threat model of many existing studies~\cite{liu2021miso,boddupalli2020redem,kim2019vehicle,boddupalli2021replace}.
}
Tu~\etal~\cite{tu2021adversarial} is the first attack specific to intermediate-fusion collaborative perception, which is an untargeted adversarial attack creating inaccurate detection bounding boxes as many as possible by perturbing feature maps in intermediate-fusion systems. However, the attack is not realistic considering the constraints of real systems, as discussed in \S\ref{sec:attack_constraints}.
\m{We propose real-world realizable attacks that challenge both early-fusion and intermediate-fusion systems.}

\noindent
\textbf{Defenses on CAV perception}.
\m{
As shown in Table~\ref{tab:related_defenses}, existing defense mechanisms are not designed for our proposed attacks thus cannot resolve them effectively.
Several Vehicle-to-Everything (V2X) communication standards~\cite{IEEE_1609_2,SAE_J3224,ETSI_TS_103324,TR_103562,hu2022gatekeeper} define security practices of network protocols (\eg access control, message integrity). They cannot block the data fabrication attacks because the attackers can modify data before wrapping it into protocol messages where the protection is enforced.
Trusted Execution Environments (TEEs)~\cite{hu2020cvshield} can potentially safeguard perception algorithms via secure hardware, but its deployment is difficult and vulnerable to side-channel attacks.
}
Against physical sensor attacks, various anomaly detection methods are proposed~\cite{liu2021stars,ranganathan2016spree,sun2020towards,alheeti2022lidar,liu2021seeing,hau2021shadow}.
For LiDAR systems especially, \tool{CARLO}~\cite{sun2020towards} detects abnormal point clouds that violate occlusion features and \tool{LIFE}~\cite{liu2021seeing} detects temporal and sensor-fusion inconsistencies.
Above defenses rely on physical rules but attackers in collaborative perception can simulate the physics to craft realistic but malicious data, as discussed in \S\ref{sec:defense_challenge}.
For connected vehicle applications, many efforts model the benign behaviors of ego/remote vehicles and detect model outliers as anomalies~\cite{liu2021miso,boddupalli2020redem,kim2019vehicle,boddupalli2021replace}. The models may involve various aspects including temporal consistency~\cite{boddupalli2020redem}, physical constraints on message delivery or vehicle control~\cite{kim2019vehicle,boddupalli2021replace}, cross-validation with local sensor~\cite{liu2021miso}, etc. 
\m{However, existing works assume the systems to share simple GPS/OBU data, making it challenging to adapt them effectively for addressing anomalies in complicated LiDAR images or feature maps.
We propose joint anomaly detection leveraging the sensing of spatial space from all connected vehicles, which enhances the spatial coverage of effective anomaly detection compared with the previous approaches.}
\mysection{Problem Definition}
\label{sec:problem}

We define the data fabrication problem in \S\ref{sec:formulation} and the threat model in \S\ref{sec:threat_model}. We emphasize the technical challenges for such new attacks compared with existing attacks in \S\ref{sec:attack_constraints}.

\mysubsection{Formulation}
\label{sec:formulation}

In a scenario where multiple vehicles jointly execute collaborative perception, the attacker aims to spoof or remove road objects (\eg vehicles, pedestrians) from designated locations in the victim's perception results.

We formulate the problem of data fabrication as an optimization problem. We denote LiDAR data at frame $i\in \mathbb{N}$ from the attacker, the victim, and other benign vehicles by $A_i$, $V_i$, and $X_i^{(j)}$, $j \in \{0, 1, \dots N\}$, respectively. LiDAR data with the same frame index will be merged on the victim side to generate perception results.
%
From Figure~\ref{fig:perception}, we denote pre-process before data sharing as $f$ and post-process after data sharing as $g$. A normal collaborative perception for the victim on frame i can be described as:
\vspace{-0.10in}
\begin{equation}
    y_i = g(f(V_i), f(A_i), f(X_i^{0}), f(X_i^{1}), ..., f(X_i^{N})).
\end{equation}

\vspace{-0.10in}
As the attacker can replace $f(A_i)$ by malicious data. For instance, the attacker can append a minor perturbation $\delta_i$ to craft malicious data as $f(A_i) + \delta_i$, which will change the original perception result from $y_i$ to $y_i'$:
\vspace{-0.10in}
\begin{equation}
    y_i' = g(f(V_i), f(A_i) + \delta_i, f(X_i^{0}), f(X_i^{1}), ..., f(X_i^{N})).
\end{equation}

\vspace{-0.10in}
Given a fitness function $I$ evaluating attack success and attack constraints $C$ restricting the perturbation, the attacker solves:
\vspace{-0.10in}
\begin{equation}
\begin{aligned}
\max_{\delta_i} I(y_i') \quad \textrm{s.t.}~C(\delta_i).
\end{aligned}
\end{equation}

\vspace{-0.10in}
\mysubsection{Threat Model}
\label{sec:threat_model}

We assume that CAVs execute collaborative perception in a Vehicle-to-Vehicle (V2V) scenario. Our results can be easily generalized to vehicle-to-infrastructure (V2I) settings by replacing one or more vehicles with edge computing devices.

We assume the attacker can physically control at least one vehicle participating in collaborative perception. This allows the attacker to gain privileges on the vehicle's software and hardware, enabling them to manipulate the sensors, tamper with the local execution of algorithms, and send arbitrary data through the network. 
In other words, attackers can directly alter the data to share, \ie, LiDAR point clouds, feature maps, and bounding boxes in early-fusion, intermediate-fusion, and late-fusion perception schemes, respectively.

\m{
We focus on early-fusion and intermediate-fusion collaboration schemes where attackers need to subtly craft complicated structured data. In terms of perception models, as the attackers locally install the perception model for joining the collaborative perception, we assume they have white-box access (i.e., model parameters). Some of our proposed attacks require no model access or only inference API.
}

Meanwhile, we assume the presence of benign vehicles which the attacker cannot invade. The assumption that the attacker would control all vehicles surrounding a victim vehicle on a busy road is deemed too impractical and financially prohibitive. We do not consider physical sensor attacks such as LiDAR spoofing~\cite{jin2022pla} and GPS spoofing~\cite{warner2003gps}. They are general threats to CAVs while we focus on new vulnerabilities brought by collaborative perception. Besides, the attacker cannot break the cryptographic protection thus cannot compromise the secure communication channels among vehicles.

\mysubsection{Attack Constraints}
\label{sec:attack_constraints}

In addition, the attacks must be realizable on real collaborative perception systems. Though Tu~\etal~\cite{tu2021adversarial} proposed a feature-perturbing attack against intermediate-fusion systems, it violates attack constraints as follows.

\noindent
\textbf{Sensor physics and definition ranges}. We require the attacker to obey basic rules in terms of the data format, otherwise it is trivial to detect the anomalies. The attackers' LiDAR point clouds should have a reasonable distribution of point density and the angle of the lasers should comply with the LiDAR configuration. In addition, the point clouds must present reasonable occlusion effects, in order to bypass anomaly detection methods based on the occlusion features~\cite{sun2020towards}.
The attackers' shared intermediate features should be within the definition ranges, avoiding absurd values.

\noindent
\textbf{Targeted attacks}. The attacker should be able to designate a target region for either spoofing or removal attacks, in order to support delicate creation of hazardous scenarios. Otherwise, the untargeted and uncontrollable attack impact as presented in Tu~\etal~\cite{tu2021adversarial} damages attack effectiveness and stealth. 

\noindent
\textbf{Real-time temporal constraints}. Collaborative perception is an asynchronous multi-agent system where each vehicle produces LiDAR images in cycles but is not synchronized in time. Figure~\ref{fig:temporal_constraint} illustrates a typical order of events in collaboration perception. To attack the victim's perception at frame i ($y_i$), the optimization of $\delta_i$ has the following constraints:

\BULLET \emph{Limited knowledge}. Optimization of $\delta_i$ must be finished before the victim's processed LiDAR data $V_i$ is generated. Therefore, attack generation cannot leverage the victim's data on the same frame. Similarly, data from other benign vehicles at frame $i$ may not be available either. The attacker can for sure rely on the shared data in previous frames from all vehicles, provided that the data transmission delay is much smaller than the LiDAR cycle. Tu~\etal~\cite{tu2021adversarial} assumes the availability of all data in the frame to attack thus it is impractical.

\BULLET \emph{Real-time attack without observable delay}. The optimization of $\delta_i$ takes time, especially when the attack involves online adversarial machine learning. To make sure $\delta_i$ is produced and transmitted before the fusion stage of the victim, the attacker can either design fast real-time attacks or optimize the perturbation before frame $i$ arrives.

\mysection{Attack Methodology}
\label{sec:attack_design}

We present realistic data fabrication attacks against various types of collaborative perception. We first introduce a general framework for real-time targeted attacks in \S\ref{sec:attack_design_overview} and elaborate on the details of ray casting attacks against early-fusion systems (\S\ref{sec:early_fusion_attack}) and adversarial attacks against intermediate-fusion systems (\S\ref{sec:intermediate_fusion_attack}). Attackers can trivially send fake bounding boxes in late-fusion systems so we omit the discussion.

\mysubsection{Zero-delay Attack Scheduling}
\label{sec:attack_design_overview}

As analyzed in \S\ref{sec:problem}, the attacks must be effective to trigger safety hazards while fast enough to satisfy real-time constraints. To satisfy both requirements, we propose an attack framework as shown in Figure~\ref{fig:attack_overview}, whose key idea is to parallelize attack generation and perception processes.

First of all, the attacker can identify the set of vehicles collaborated with the victim vehicle and align frame indices of their shared sensor data based on timestamps. The attack generation module is triggered on each LiDAR cycle.
It first tracks the target region: (1) for object spoofing, the trajectory of the object to spoof is predefined; (2) for object removal, the attacker needs a simple object detection algorithm to localize the target object to remove. Then it optimizes the malicious perturbation that can be used to attack the victim's perception at the current frame. Note that the optimized perturbation is generated overtime and cannot be used to attack due to the real-time constraints (\S\ref{sec:attack_constraints}). We need to transform the perturbation into one that has a similar attack impact on the next frame.
In this way, the perturbation is ready to apply when the next frame arrives, introducing no additional delay to the original collaborative perception pipeline. As the attack generation occurs one frame in advance, it affords the attacker up to one LiDAR cycle time to complete the optimization.

The optimization and transformation of the perturbation highly depend on the configuration of the collaborative system and will be discussed in later sections.


\begin{figure*}[!htb]
\minipage{0.70\textwidth}
    \includegraphics[width=\textwidth,height=0.55\textwidth]{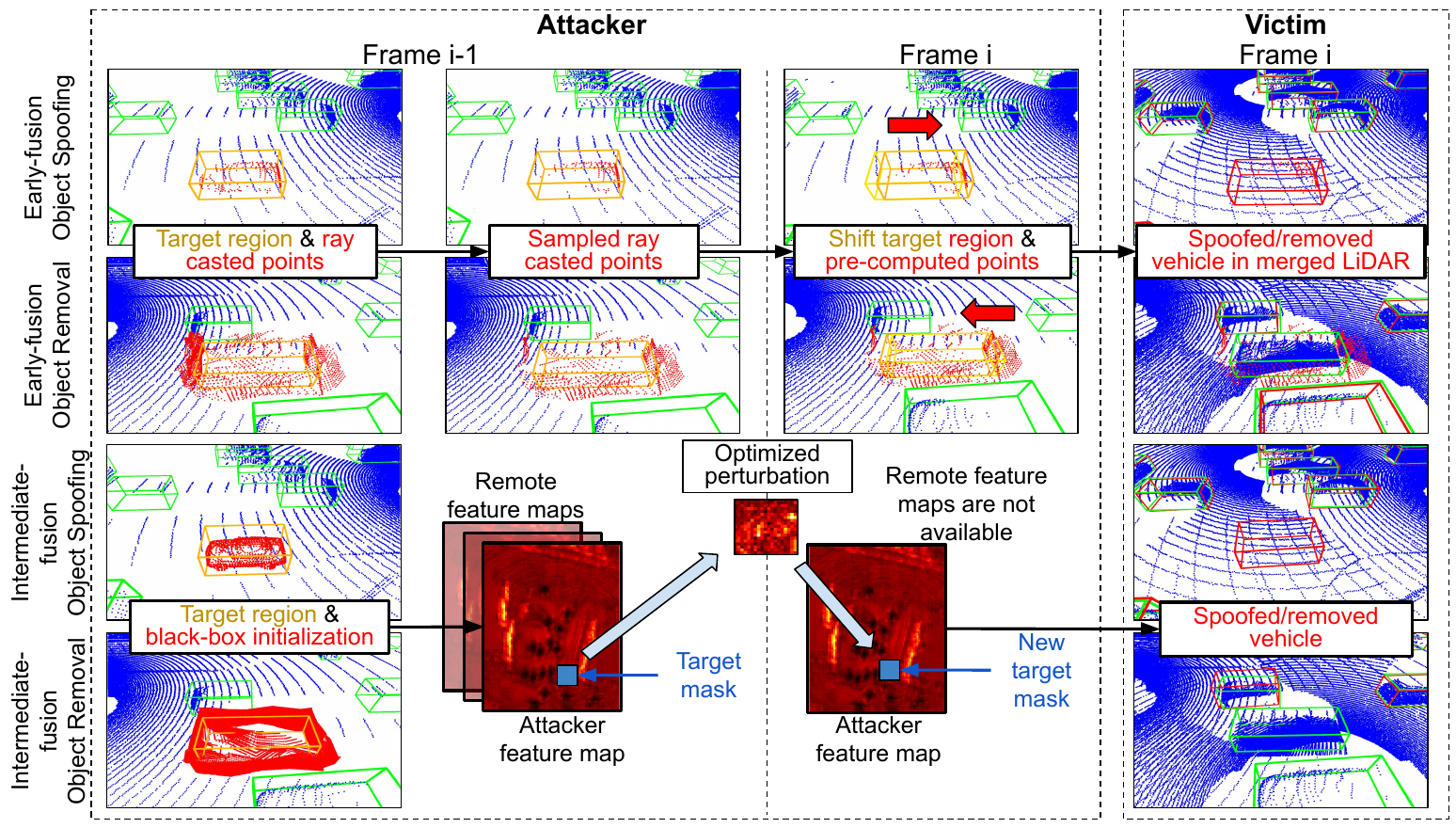}
    \vspace{-0.25in}
    \caption{Demonstration of collaborative perception attacks. The orange box is the target region, the green boxes are ground-truth labels, and the red boxes are predicted objects.}
    \vspace{-0.15in}
    \label{fig:attack_demo}
\endminipage\hfill
\minipage{0.28\textwidth}
    \includegraphics[width=\textwidth,height=1.4\textwidth]{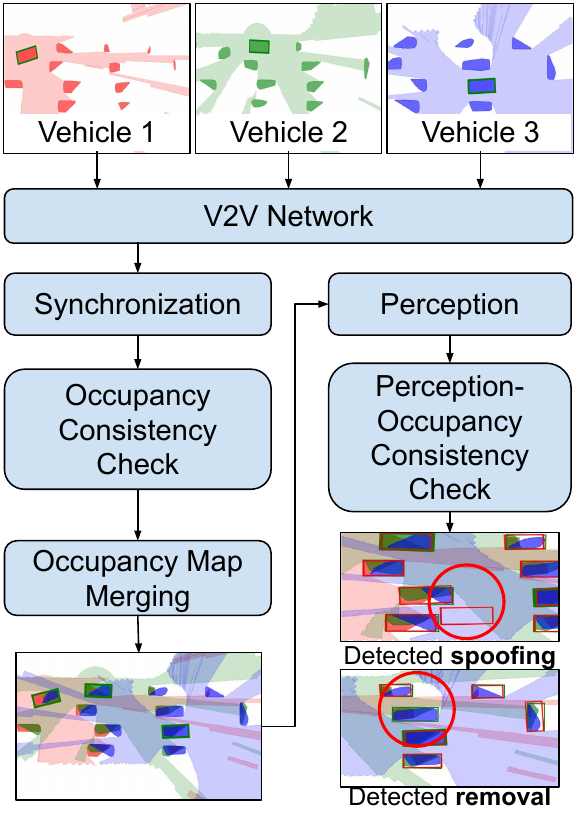}
    \vspace{-0.25in}
    \caption{Workflow of \system's anomaly detection.}
    \vspace{-0.15in}
    \label{fig:system_overview}
\endminipage
\end{figure*}

\mysubsection{Black-box Ray Casting Attack}
\label{sec:early_fusion_attack}

In early-fusion collaborative systems, CAVs share LiDAR point clouds. Thus, the attacker will perturb the location of LiDAR points directly but must obey the physical rules of LiDAR sensors as mentioned in \S\ref{sec:attack_constraints}. Note that a white-box adversarial attack~\cite{cao2019adversarial} is not applicable because (1) most perception models involve non-differential pre-processing and (2) even if the gradient can be approximated, the heavy computation can hardly achieve real-time attacks.

\noindent
\textbf{Insights}.
First, we find that a higher point density on the object surface leads to more successful detection. Mainstream 3D object detection models learn spatial features from voxelized point groups (\S\ref{sec:background}). It is therefore natural that a higher point density strengthens the learned feature toward object classes.
Second, a higher coverage on object surfaces also contributes to better detection, as the shape features of objects become more explicit. This is also one of the key benefits of collaborative perception, as multi-view LiDAR data allows for a more comprehensive perception of objects.  
Given the two insights, the object spoofing attack aims to spoof denser LiDAR points of objects and cover a larger surface area of the object. The goal of the object removal attack is to obscure the surface of the original object as thoroughly as possible. We confirm the insights in our ablation study (\S\ref{sec:attack_ablation}).

\begin{algorithm}[t]
\scriptsize
\KwIn{A target region $y_t$, LiDAR image $X$, an 3D object model $S$ (generated offline using an initial model $S_0$ and an attack dataset $D$).}
\KwOut{Fake LiDAR image $X_a$.}
    \SetKwFunction{FMain}{BlackboxAttack}
    \SetKwProg{Fn}{Function}{:}{}
    \Fn{\FMain{$X$, $y_t$, $S$}}{
        $S_t \gets \texttt{Transform}(S, y_t)$\;
        $X_{r} \gets \texttt{NonOcclusionRayCasting}(X, S_t)$\;
        $X_a \gets \texttt{PointSampling}(X_r, S_t)$\;
    }
    \SetKwFunction{FMain}{AdversarialShape}
    \SetKwProg{Fn}{Function}{:}{}
    \Fn{\FMain{$S_0$, $D$}}{
        $S \gets S_0;$ \Comment{Only for object removal.}\\
        \For {Iteration 1\dots K} {
            \For {$X^{(i)}, y_t^{(i)} \in D$} {
                $X_a^{(i)} \gets \texttt{BlackboxAttack}(X^{(i)}, y_t^{(i)}, S)$\;
                $Y^{(i)} \gets \texttt{Perception}(X_a^{(i)})$\;
                Optimize $S$ by maximizing $\sum_{y\in Y^{(i)}} \texttt{IoU}(y, y_t^{(i)}) \cdot log(y_\sigma)$\; \Comment{$y_\sigma$ is the confidence score.}
            }
        }
    }
 \caption{\m{Black-box ray-casting attacks.}}
 \label{alg:black_box}
\end{algorithm}

\noindent
\textbf{Attack methods}
The attacker pretends that an object is spoofed or removed and reconstructs the LiDAR point cloud via ray casting techniques. The traced rays follow the physical laws of the original lasers so the reconstructed point cloud is realistic. 
\m{The spoofing attack requires no model access while the removal attack requires the model's inference API.}
The attack is demonstrated in Figure~\ref{fig:attack_demo} and Algorithm~\ref{alg:black_box}.

\emph{Preparation of 3D object model}. The attacker first constructs a 3D model (\eg a triangle mesh) of the object they wish to fabricate. In later attack steps, we will place the 3D model in the target region and cast malicious points on its surfaces.
For object spoofing, the model can represent a real object such as a car. For object removal, we optimize a universal adversarial shape offline as the model. We initialize a cuboid triangle mesh and use a black-box genetic algorithm to optimize the perturbation on mesh vertices. 
\m{As shown in Algorithm~\ref{alg:black_box} (\texttt{AdversarialShape}), in each iteration, we launch the object removal attack on a dataset of attack cases and optimize the object model to maximize a fitness score representing the success of attacks (i.e., minimizing the confidence of detection proposals in the target region).} A detailed explanation is in Appendix~\ref{sec:adv_shape}.


\emph{Non-occlusion ray casting}. We set up a ray casting scenario where the 3D model is placed at the designated location and the rays are lasers in the attacker's LiDAR image.
\m{
Though the predefined object models have fixed sizes, we will dynamically adjust size, location, and orientation of them to fit the target region during the scenario creation (\texttt{Transform} in Algorithm~\ref{alg:black_box}), making the object models universal for various attack situations.
}
The ray casting algorithm calculates the points of intersection between the rays and the 3D model. To maximize point density on the target object, the ray casting is customized to ignore occlusion effects, ensuring that each ray is not blocked and goes through model surfaces to leave multiple intersection points.

\emph{Point sampling}. We resolve the occlusion violations by sampling one intersection point per ray. Specifically, for each ray with one or more intersection points with the 3D model, its original LiDAR point is replaced by one of the intersection points. The selection of intersection points is through customizable weighted random sampling. In our implementation, intersection points closer to benign vehicles have a higher probability of being selected.
In this way, spoofed fake points tend to have a higher density close to benign points, increasing the chance of obscuring the original point distribution. Also, the randomness ensures high coverage on object surfaces.
More details are presented in Appendix~\ref{sec:point_sampling}.

\noindent
\textbf{Attack transformation}. To transform the attack into a future frame, we need to record the modified LiDAR points and corresponding ray angles. When the next frame is produced, the attacker removes points with the same ray angles, transforms recorded points to the new target region, and appends the transformed points. Since two frames have a minor time interval (100 ms), the transformation preserves physical laws.

\noindent
\textbf{Time constraint}. The attack generation can start when the attacker's LiDAR image is produced. Though \emph{point sampling} requires the locations of remote LiDARs, they can be predicted using simple linear velocity estimation. The ray casting should be done within one LiDAR cycle.

\mysubsection{White-box Online Adversarial Attack}
\label{sec:intermediate_fusion_attack}

Intermediate-fusion systems require CAVs to exchange feature maps, the intermediate result of neural network processing. Such systems are immune to the black-box ray casting attacks (\S\ref{sec:early_fusion_attack}) because the presence of benign feature maps will drop the attack success rate significantly, as demonstrated later in our experiments (\S\ref{sec:attack_ablation}).
Adversarial machine learning, on the other hand, is able to generate adversarial feature maps. \m{The attack assumes that the attacker has white-box knowledge of perception models.}

\noindent
\textbf{Insights}. We optimize a perturbation on the attacker's feature map by performing a backward pass in each LiDAR cycle and reusing the perturbation over frames as an online attack, similar to Tu et. al.~\cite{tu2020physically}. We introduce two new ideas to achieve realistic real-time targeted attacks.

First, we initialize the perturbation using results from black-box ray casting attacks, making the initial perturbation vector closer to the optimal choice. This step is crucial for achieving real-time attacks as it significantly reduces the number of optimization iterations required.

Second, to restrict attack impact to a specific region, we mask the feature map. This is based on the fact that convolution networks preserve the relationship between feature map indices and real-world locations~\cite{lang2019pointpillars}. Another conventional approach to enforce spatial constraints is to add a regularization term to the loss function (\eg penalize detection errors in non-attack regions). However, this requires multiple iterations to converge, making it unsuitable for real-time attacks.

\noindent
\textbf{Attack methods}.
The attacker optimizes a perturbation on their feature map over continuous frames. For each frame, the attacker spoof/remove objects in the point cloud first as initialization, then updates the latest perturbation map through an iteration of projected gradient descent (PGD). As the target region moves, the perturbation is re-indexed accordingly in each cycle. The key steps are demonstrated in Figure~\ref{fig:attack_demo}.

\emph{Black-box initialization}.
The attacker starts by modifying the raw point cloud. Unlike the ray casting attack in \S\ref{sec:early_fusion_attack}, there is no restriction on this modification in terms of the physical laws. Therefore, the attacker tends to inject high-density high-coverage LiDAR points representing the 3D models mentioned in \S\ref{sec:early_fusion_attack}, which can be prepared offline.

\emph{Feature map masking}.
We make the assumption that each feature map index is associated with a voxel/pillar in the 3D real-world coordinate system. Given the target region, we extend the region by a fine-tuned parameter and extract corresponding feature indices. The masking operation ensures that only features with the selected indices are perturbed. If the index mapping is not explicit, it can be approximated by comparing the feature map before and after the black-box initialization and identifying the indices where the feature values have been altered.

\emph{Loss objective}.
The optimization objective is to increase/decrease the score of the bounding box proposal on the labeled attack region, for spoofing/removing objects. We define the objective function as Equation~\ref{eq:objective}, where $Z'$ denotes the set of bounding box proposals after the perturbation, $z'_\sigma$ is the score associated with the proposal $z'$, and $z_{t}$ represents the target region to attack. The objective function maximizes/minimizes the confidence score of proposals overlapping with the target.
\vspace{-0.06in}
\begin{equation}
\label{eq:objective}
\begin{gathered}
l_{spoof}(Z') =  \sum_{z' \in Z'} \textproc{IoU}(z', z_{t}) \cdot \log(1 - z'_{\sigma}) \\
l_{remove}(Z') =  - \sum_{z' \in Z'} \textproc{IoU}(z', z_{t}) \cdot \log(1 - z'_{\sigma})
\end{gathered}
\end{equation}

\vspace{-0.10in}
\emph{Constraints on perturbation}. We clip the perturbation by restricting feature values to their normal range, which is measured on a set of non-attack test cases. As feature values do not explicitly deliver spatial semantics that can be used for anomaly detection, there is no need to restrict feature perturbation to minor thresholds.

\noindent
\textbf{Attack transformation}. Given the centers of target regions between two consecutive frames, one can get corresponding feature map indices $(x_0, y_0)$ and $(x_1, y_1)$ respectively. Then each index $(i, j)$ in the feature map is mapped to $(i-x_0+x_1,j-y_0+y_1)$.

\noindent
\textbf{Time constraint}. The PGD optimization needs feature maps shared from as many vehicles that cooperate with the victim. Assuming all benign vehicles continuously broadcast and process feature maps at a frequency equal to the LiDAR cycle ($T$) and the transmission delay is below a threshold $t_T$, the optimization must be done within $T-2t_T$.

\mysection{Anomaly Detection}
\label{sec:defense_design}

We propose \system, a \underline{C}ollaborative \underline{A}nomaly \underline{D}etection system to mitigate the security threats presented in \S\ref{sec:attack_design}. We enumerate the design challenges in \S\ref{sec:defense_challenge}. In \S\ref{sec:defense_system}, we outline our system, followed by the details of key components in \S\ref{sec:occupancy_map} and \S\ref{sec:consist_checks}.

\mysubsection{Challenges}
\label{sec:defense_challenge}

As discussed in \S\ref{sec:background}, existing defense mechanisms~\cite{sun2020towards,liu2021seeing,ambrosin2019design} mainly focus on finding temporal or spatial inconsistencies but they cannot handle attackers who can generating fake data that conform with physics laws.
We propose a cross-agent consistency check where all benign vehicles exchange evidence of anomalies to reveal adversarial behaviors jointly. To ensure the effectiveness, robustness, and generality of the proposed method, we have to overcome the following challenges. 

\emph{Affordable bandwidth and computation cost}.
Collaborative perception systems must finish a perception cycle within a hard deadline (\eg 100 ms~\cite{lin2018architectural}). Therefore, \system should only share minimal, essential data to save bandwidth and distribute data processing on different vehicles to minimize latency. Our method only shares small-sized metadata.

\emph{Detection of stealthy attacks}.
As the attacks may inject malicious data into a specific small region in 3D space, fine-grained anomaly detection is required. For instance, spoofing a ghost vehicle affects a region of approximately 10~$m^2$ while the perception range is over 4,000~$m^2$. \system uses fine-grained occupancy maps to precisely reveal abnormal regions.

\emph{Robustness to benign errors}.
LiDAR data captured by different vehicles have slight differences in timestamps~\cite{wang2020v2vnet,shi2022vips}. 
\system leverages motion estimation and prediction to synchronize occupancy maps. Localization error is another potential source of faults. As nowadays vehicle localization achieves an accuracy of less than 0.1~m~\cite{oxts}, \system can tolerate minor errors with proper threshold parameters.

\mysubsection{System Overview}
\label{sec:defense_system}

\system is a system deployed on CAVs against data fabrication during collaborative perception. As shown in Figure~\ref{fig:system_overview}, besides the original perception pipeline, CAVs are required to perform anomaly detection tasks in parallel.


When a local LiDAR image is produced, each vehicle generates an occupancy map that labels on-road objects, free-to-drive regions, and invisible regions in the 2D space. Then the occupancy map is broadcast via a V2V wireless network. The occupancy map is represented in fine-grained polygons, balancing precision and transmission overhead. In addition, motion information of on-road objects is attached for synchronizing occupancy maps from different vehicles.

After collecting occupancy maps from other vehicles, each vehicle launches two consistency checks.
\emph{Occupancy consistency check} reveals inconsistencies of occupancy maps, \eg one region identified as free and occupied by two different vehicles indicates that one of the participants is faulty or malicious. Occupancy maps are then merged into one, with inconsistent regions marked as unknown.
\emph{Perception-occupancy consistency check} then ensures the results of collaborative perception are consistent with the merged occupancy map - bounding boxes should overlap with occupied regions instead of free regions; on-road occupied regions should be detected in at least one bounding box.
Even though attackers can launch strong stealthy attacks and fake occupancy maps, the attack impact is always reflected by perception results and can be revealed as malicious by benign occupancy maps.  


\mysubsection{Occupancy Map}
\label{sec:occupancy_map}

The occupancy map generation involves three steps: point segmentation, space segmentation, and motion estimation.

\emph{Point segmentation}. First, we eliminate less useful background points that are not on the road using HD maps provided by autonomous driving systems~\cite{apollo,autoware}. Then, we apply ground fitting algorithms (\eg RANSAC~\cite{fischler1981random} to detect the ground plane and remove LiDAR points on it. By clustering the remaining points based on point density, we can identify all non-ground objects on the road, with each cluster representing a unique on-road object. The method has been proven to be effective in prior research~\cite{douillard2011segmentation, wu2018squeezeseg, zermas2017fast}.

\emph{Space segmentation}. After identifying on-road objects, we generate a fine-grained representation of 2D space occupancy, which classifies the 2D space into three categories: \emph{free}, \emph{occupied}, and \emph{unknown}. 
(1) Occupied regions are the convex hulls~\cite{avis1995good} of the object clusters.
(2) Free regions represent the region surrounded by only ground points. We evenly divide the 2D space into equal sectors whose vertex is the LiDAR sensor location. The number of sectors can be adjusted for different levels of granularity. In each sector, we measure the distance from the LiDAR to the closest non-ground point and label the region within the distance as a free region. A basic implementation of free regions is described above, while we introduce an optimized implementation in Appendix~\ref{sec:space_seg}.
(3) The remaining region is classified as unknown due to occlusion or the limited range of LiDAR sensors. Since the accuracy of segmentation and clustering drops as LiDAR points get sparser, in the implementation, we define a 2D space as unknown if its distance to the LiDAR sensor exceeds a threshold (\eg 50~m). Unlike conventional grid-based occupancy maps~\cite{kim2017probabilistic, qiu2021autocast, li2014multivehicle}, our occupancy map divides regions using polygon representation. Our approach offers two advantages over grid representation: (1) polygons can more precisely depict arbitrary shapes; (2) by adjusting the outline smoothing factor, polygon representation provides greater flexibility to strike an optimal balance between precision and size.

\emph{Motion estimation}.
First, each CAV executes a multi-object tracking (MOT) process on object point clusters. Inspired by \tool{AB3DMOT}~\cite{weng20203d}, a baseline solution of MOT, we assign an affinity score to each object pair between two consecutive frames. This affinity score indicates the level of similarity considering factors such as distance and point density. Using the scores, MOT algorithms can match the same object across frames.
Second, given two point clusters that refer to the same object but on two consecutive frames, we use point cloud registration to derive a transformation matrix between them.
Formally, if two object clusters with timestamp $t'$ and $t$ ($t - t' \approx T$ where $T$ is LiDAR cycle time) are denoted as $X_{t'}$ and $X_t$ respectively, the transformation matrix $T_t$ satisfies $X_t = T_t \cdot X_{t'}$. We then standardize the matrix to \emph{motion per time unit} - divide translation and rotation extracted from $T_t$ by the time gap $t-t'$ and reconstruct the matrix as $T_e$. We define this operation as \textproc{Scale}: $T_e = \textproc{Scale}(T_t, \frac{1}{t-t'})$.
$T_e$ represents the latest motion of the specific object and is attached to the corresponding occupied region in the occupancy map.
Also, the maps should be transformed into a global coordination system as a consensus of all CAVs.


\mysubsection{Consistency Checks}
\label{sec:consist_checks}

The processes of consistency checks are triggered simultaneously with the data fusion, involving occupancy map synchronization, occupancy consistency checks, and perception-occupancy consistency checks.

\emph{Occupancy map synchronization}. After receiving a set of occupancy maps with slightly different timestamps, each vehicle aims to synchronize all maps to the timestamp of the latest local LiDAR image.
For each on-road occupied region in each occupancy map (except the local map), we first calculate its time gap to the target timestamp, denoted by $\Delta_t$. We then transform the occupied region by applying the transformation $\textproc{Scale}(T_e, \Delta_t)$, where $T_e$ is the corresponding motion per time unit. After moving all occupied regions, we post-process the occupancy map by excluding new occupied regions from the original free regions to resolve conflicts.
In this way, all occupancy maps can be directly merged as they have been synchronized spatially and temporally.
Formally, we denote synchronized occupancy maps by $M^{(i)}=(S_O^{(i)}, S_F^{(i)})$ where $i\in \{0,1,\dots,N\}$ denotes vehicle IDs ($t=0$ denotes the ego vehicle) and $S_O$/$S_F$ denotes occupied/free regions.

\emph{Occupancy consistency check} reveals inconsistencies among synchronized occupancy maps. A region is considered conflicted if it is identified as occupied by one vehicle and free by another.
We can define conflicted regions as

\vspace{-0.10in}
\begin{equation}
    \epsilon_{occ} = \bigcup_{i,j\in 0\dots N} S_O^{(i)} \cap S_F^{(j)}.
\end{equation}

\vspace{-0.10in}
Considering the inevitable imperfection of synchronization, in the implementation, \system will ignore conflict regions whose area is below a threshold (\ie $\sigma_{occ}$). Alerts are raised indicating the uncertain risks on conflicted regions.

Next, each vehicle generates one consistent occupancy map by merging available occupancy maps and dropping conflicted regions. Particularly, the occupancy map produced by the ego vehicle is trusted and retained in the merged map, \m{unless sensors of the ego vehicle is detected as compromised by existing detection of LiDAR spoofing~\cite{sun2020towards,liu2021seeing}}. The new occupancy map $M'=(S_O', S_F')$ is generated as:

\vspace{-0.10in}
\begin{equation}
\begin{gathered}
S_O' = S_O^{(0)} \cup (\bigcup_{i=1\dots N} S_O^{(i)} - \epsilon_{occ} - S_F^{(0)}) \\
S_F' = S_F^{(0)} \cup (\bigcup_{i=1\dots N} S_F^{(i)} - \epsilon_{occ} - S_O^{(0)})
\end{gathered}
\end{equation}
\vspace{-0.10in}

\vspace{-0.03in}
\emph{Perception-occupancy consistency check} aims to reveal inconsistencies between the perception results and the merged occupancy map based on two rules.
First, free regions should have overlap with predicted object bounding boxes. According to LiDAR sensor physics, objects on the road, if observable, always leave LiDAR points above the ground and should be clustered as occupied regions. This rule can counter object spoofing attacks, as attackers may spoof fake objects in free regions perceived by benign vehicles.
Second, occupied regions should be within predicted bounding boxes. Similarly, point clusters on roads are potential obstacles and should be detected to avoid a collision. It serves as a countermeasure against object removal attacks where attackers make real objects undetectable.
By checking the two rules, alerts are raised on conflicted regions, similarly filtered by a threshold of area (\ie $\sigma_{spoof}$ and $\sigma_{remove}$). Formally, if we denote predicted bounding boxes as $Y$, alerted regions include:

\vspace{-0.10in}
\begin{equation}
\label{eq:detection}
\epsilon_{spoof} = \bigcup_{y \in Y} y \cap S_F' \quad
\epsilon_{remove} = \bigcup_{s_O' \in S_O'} s_O' - Y
\end{equation}
\vspace{-0.10in}



\vspace{-0.20in}
\subsection{Limitations}

\m{
\system is a mitigation other than elimination of our proposed attacks. 
First, \system cannot work in certain extreme scenarios. The detection could be successful only when at least a benign CAV observes the attacked region. Otherwise, the attacked region is an occluded region for all benign CAVs thus no conflict will appear in Equation~\ref{eq:detection}. 
Second, \system detects but may not resolve the anomalies. Though the system may identify the possible attackers via majority voting, it is limited in effectiveness if benign CAVs do not dominate the road.}

\mysection{Evaluation}

We introduce our dataset creation in \S\ref{sec:dataset} and the implementation details in \S\ref{sec:implementation}. Then, we present a comprehensive evaluation of the proposed attacks and defenses in \S\ref{sec:attack_eval} and \S\ref{sec:defense_eval}.

\mysubsection{Data Collection}
\label{sec:dataset}

\noindent
\textbf{Adv-OPV2V}. \tool{OPV2V}~\cite{xu2022opv2v} is a benchmark dataset for collaborative perception algorithms, with data collected from a combination of simulators, \tool{CARLA}~\cite{carla} and \tool{SUMO}~\cite{sumo}. We generate \tool{Adv-OPV2V} from \tool{OPV2V}, as a benchmark for testing collaborative perception attacks and defenses. We select 300 scenarios for object spoofing and removal attacks respectively. Each scenario features 10 consecutive frames and 3 to 5 CAVs among which one attacker and one victim are designated. Each scenario also has predefined attack targets, such as a trajectory of a ghost vehicle for object spoofing or a trajectory of an existing vehicle for object removal. To ensure the real-world impact of the attacks, we limit the distance between the victim and the target to less than 30~m.

\noindent
\textbf{Adv-\testbed}. We create a real-world multi-vehicle collaborative perception dataset using testbed \testbed~\cite{mcity}, which is a real-world mock city for testing CAV applications. On real roads, we deploy 3 Lincoln MKZ vehicles as CAVs, which are equipped with OxTS RT3000v3 GPS, Velodyne VLP-32C LiDAR, and Cohda MK6C OBU as a C-V2X receiver. We also deploy several other vehicles as perception targets.
We create 8 attack scenarios that contain potential safety hazards, with 4 for object spoofing and 4 for object removal.
We collect LiDAR, GPS, and C-V2X network traces from all CAVs to allow for emulation of collaborative perception.


\mysubsection{Implementation}
\label{sec:implementation}

\noindent
\textbf{Collaborative perception models}. For \tool{Adv-OPV2V}, we utilize pre-trained models provided by \tool{OPV2V}, which employ naive point cloud merging in early-fusion and attentive learning in intermediate-fusion.
\m{
In early-fusion methods, the point clouds are naively concatenated together. In intermediate-fusion methods, the fusion is defined by the models.
}
For \tool{Adv-\testbed}, we augment the \tool{OPV2V} training data to approximate the LiDAR images collected in the testbed \testbed and fine-tune the pre-trained models.
\m{
During the training of models, an uniform noise of at most 0.2~m or 0.2$^\circ$ is injected to vehicle locations or rotations respectively, in order to better tolerate localization/synchronization errors in real scenarios, following the previous work~\cite{wang2020v2vnet,lu2022robust,xu2022v2x}.}

\noindent
\textbf{Attacks} are implemented in 4,874 lines of code (LOC) in Python. The adversarial shape generation is based on a classic genetic algorithm with a population size of 10 and for 5 generations. Adversarial attacks are based on \tool{Torch}. We fine-tune the learning rate to 1 and optimize for a maximum of 25 iterations. The perturbation of feature maps is restricted in a 5~m$\times$5~m square centered by the target location.

\noindent
\textbf{Anomaly detection} is implemented in 1,629 LOC in Python, which uses polygon operations in \tool{shapely} and implementation of RANSAC and DBSCAN from \tool{Open3D}. The system parameters (\ie $\sigma_{occ}$, $\sigma_{spoof}$, $\sigma_{remove}$) are not fixed but evaluated through the receiver operating characteristic (ROC) curve. 

\noindent
\textbf{In-vehicle execution environment}. To demonstrate system deployment on real vehicles, we implement a collaborative perception framework based on Robot Operating System (ROS), consisting of 3,154 LOC in C++ responsible for V2V communication and basic sensor data processing. Our implementation of attacks and anomaly detection can be plugged into the framework as ROS nodes. For performance measurement, we use an in-vehicle machine with an Intel Xeon Silver 4110 CPU and an Nvidia RTX 2080 Ti GPU.

Our implementation is open source at \url{https://github.com/zqzqz/AdvCollaborativePerception}.

\mysubsection{Evaluation of Attacks}
\label{sec:attack_eval}


We present our attack results in \S\ref{sec:attack_results}. We further analyze the impacting factors in attacks (\S\ref{sec:attack_factors}) and present an ablation study (\S\ref{sec:attack_ablation}). We realize attacks in the testbed \testbed, evaluate the overhead (\S\ref{sec:attack_performance}) and conduct case studies (\S\ref{sec:attack_testbed}).





\mysubsubsection{Attack Results}
\label{sec:attack_results}

\begin{table}[t]
  \scriptsize
  \caption{Performance of attacks and defenses on \tool{Adv-OPV2V}.}
  \vspace{-0.04in}
  \label{tab:attack_results}
  \setlength{\tabcolsep}{3.66pt}
  \begin{tabular}{| c | c | c | c | c | c | c | c | c |}
    \noalign{\global\arrayrulewidth1pt}\hline\noalign{\global\arrayrulewidth0.4pt}
    Attack setting: & \multicolumn{4}{c|}{Attack results} & \multicolumn{3}{c|}{Defense results} \\
    \cline{2-8}
    Method-Fusion-Goal & Succ. & IoU & Score & $\Delta$AP & Succ. & TPR & FPR \\
    \noalign{\global\arrayrulewidth1pt}\hline\noalign{\global\arrayrulewidth0.4pt}
    \m{\cite{tu2021adversarial}-Int.-Spoof} & \m{21.7\%} & \m{0.01} & \m{0.06} & \m{-62.8\%} & \m{100\%} & \m{34.0\%} & \m{10.3\%} \\
    \m{\cite{tu2021adversarial}-Int.-Remove} & \m{14.0\%} & \m{0.47} & \m{0.34} & \m{-61.8\%} & \m{100\%} & \m{39.7\%} & \m{7.6\%} \\
    RC-Early-Spoof & 86.0\% & 0.55 & 0.38 & -0.4\% & 83.8\% & 80.9\% & 2.0\% \\
    RC-Early-Remove & 87.3\% & 0.07 & 0.03 & -0.5\% & 81.2\% & 38.0\% & 5.6\% \\
    Adv.-Int.-Spoof & 90.0\% & 0.46 & 0.71 & -2.0\% & 83.4\% & 80.1\% & 2.0\% \\
    Adv.-Int.-Remove & 99.3\% & 0.02 & 0.01 & -3.9\% & 83.6\% & 42.5\% & 2.2\% \\
    Naive-Late-Spoof & 98.7\% & 0.96 & 0.99 & 0 & 80.8\% & 84.8\% & 2.7\% \\
    Naive-Late-Remove & 0.3\% & 0.78 & 0.53 & 0 & - & - & - \\
    \noalign{\global\arrayrulewidth1pt}\hline\noalign{\global\arrayrulewidth0.4pt}
  \end{tabular}
  {\notsotiny \textbf{Notes:} \emph{Int.} - intermediate-fusion. \emph{RC} - ray casting. \emph{Adv.} - adversarial attack. \emph{Succ.} - success rate.}
  \vspace{-0.05in}
\end{table}

To evaluate attack effectiveness, we launch each proposed attack on 300 attack scenarios in \tool{Adv-OPV2V} against baseline perception models using \tool{PointPillars}~\cite{lang2019pointpillars} as the backbone. Attack results are listed in Table~\ref{tab:attack_results}.
In each attack scenario, we identify the best predicted bounding box having the largest Intersection over Union (IoU) with the target region.
A spoofing attack is considered successful if the IoU is greater than zero while a removal attack is considered successful if the IoU is zero.
For late-fusion systems, object spoofing is trivial to reach almost 100\% success rate while object removal is hard as long as one benign vehicle observes the object.
Our proposed attacks against early/intermediate-fusion are generally successful with a success rate above 86\%.

\begin{figure} 
    \centering
    \includegraphics[width=0.47\textwidth]{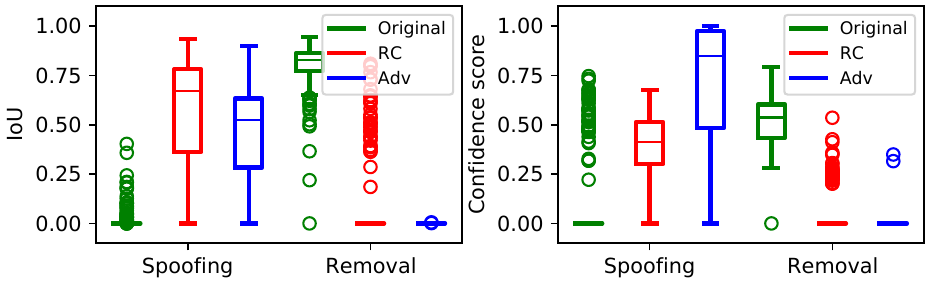}
    \vspace{-0.06in}
    \caption{IoU/confidence on target region under \m{the prior attack}, our ray casting (RC) and adversarial (Adv) attacks.}
    \vspace{-0.025in}
    \label{fig:iou_score}
\end{figure}

\begin{figure} 
    \centering
    \includegraphics[width=0.47\textwidth]{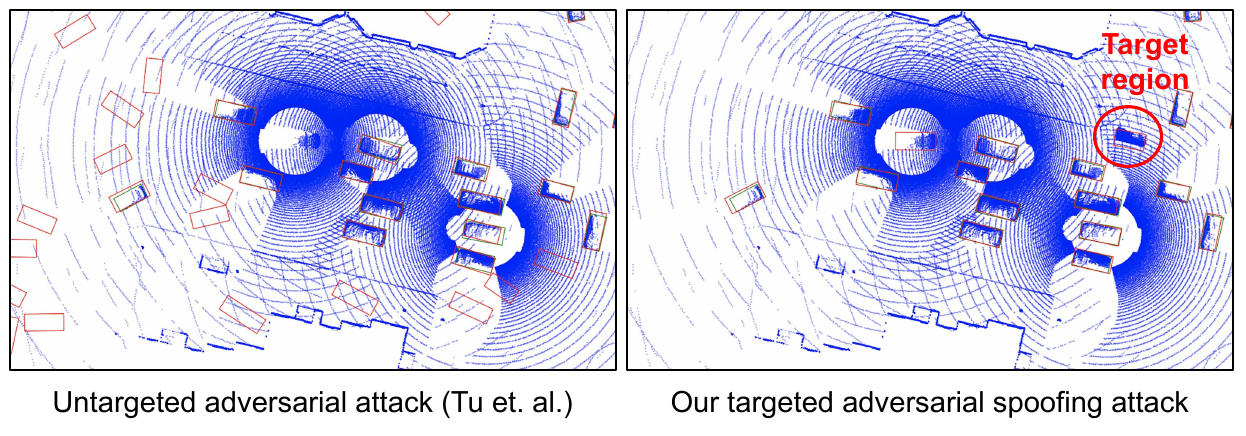}
    \vspace{-0.10in}
    \caption{\m{Different stealth of untargeted/targeted attacks.}}
    \vspace{-0.10in}
    \label{fig:stealth_demo}
\end{figure}

In addition, we illustrate the change of IoU and confidence score on target regions in Figure~\ref{fig:iou_score}. We observe that attacks make a significant change in the two metrics. For spoofing attacks, the early-fusion ray casting attack achieves a larger IoU meaning more accurate spoofed bounding boxes while the intermediate-fusion adversarial attack pushes the confidence score to extremely high (> 0.8).
The result indicates that attacker is easier to launch sophisticated attacks against early-fusion systems since attackers can directly manipulate the subtle spatial features - LiDAR points.
The intermediate-fusion system enforces fewer constraints on malicious perturbation thus the upper bound of the attack impact is higher.
The change of Average Precision ($\Delta AP$) on non-attack regions is minor, which means all attacks focus on perturbing the target region. However, $\Delta AP$ of intermediate-fusion attacks is higher because the perturbation on feature maps inevitably propagates to a larger region through convolution layers. 

We also reproduced the prior attack proposed by Tu \etal~\cite{tu2021adversarial} 
\m{We use the loss function and attack parameters from the paper and the constraint of perturbation the same as our adversarial attack. We also allow the unrealistic attack constraints as discussed in \S\ref{sec:threat_model}.
Attack results on \tool{Adv-OPV2V} are shown in Table~\ref{tab:attack_results} and Figure~\ref{fig:iou_score}.
The prior attack is successful in its attack goal to inject as many false perception bounding boxes as possible, by injecting on average 56.2 FPs and 3.6 FNs in each LiDAR frame. The overwhelming FPs drop $AP$ to nearly 1\%.
However, when attacking a certain region, it only yields 14\%-22\% success rate because the attack is untargeted fundamentally.
The major problem of the untargeted approach is stealth of attacks. The untargeted attack generates a significant number of abnormal bounding boxes that are out of the road or heading away from the lane direction, as shown in Figure~\ref{fig:stealth_demo}.}
The uncontrollable attack impact can be easily recognized by either humans or automatic anomaly detection. 

\mysubsubsection{Impacting Factors}
\label{sec:attack_factors}

\begin{figure}
    \centering
    \begin{subfigure}[b]{0.48\textwidth}
        \centering
        \includegraphics[width=\textwidth]{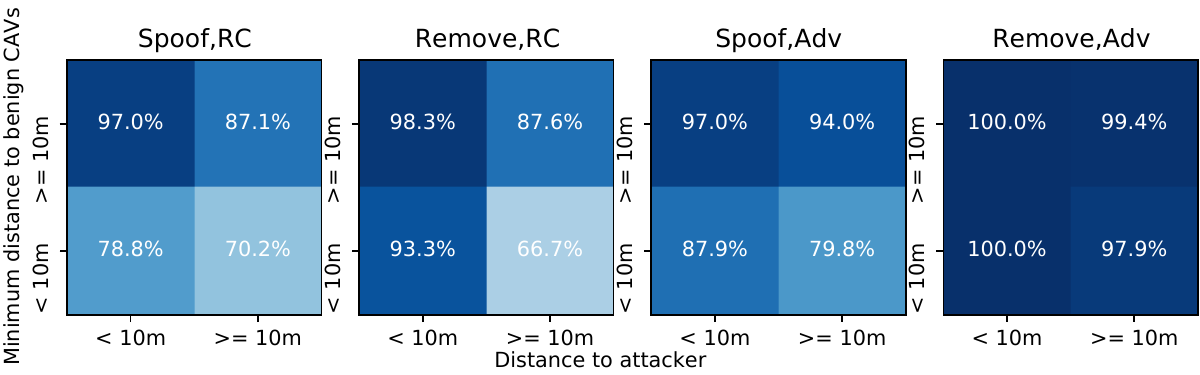}
    \end{subfigure}\hfill
    \begin{subfigure}[b]{0.48\textwidth}
        \centering
        \includegraphics[width=\textwidth]{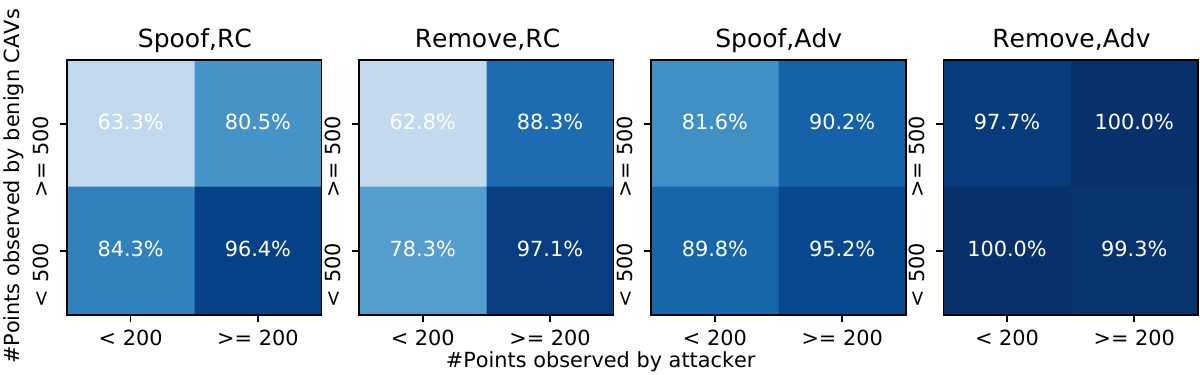}
    \end{subfigure}
    \vspace{-0.22in}
    \caption{\m{Attack success rate w.r.t. target visibility.}}
    \vspace{-0.08in}
    \label{fig:factor_visibility}
\end{figure}

\noindent
\textbf{Visibility of the target region}. 
We hypothesize that the attack is more successful when the target region is clearly visible to the target but not benign CAVs.
\m{
Intuitively, the target is more visible if it is closer to the LiDAR or there are more LiDAR points on it.
To validate the hypothesis, we draw relationship between attack success rate and the two metrics in Figure~\ref{fig:factor_visibility}.
The result shows that the attack is more successful when the attacker is closer to the target while benign CAVs are further away, or the attacker has more LiDAR points in the target region while benign vehicles have fewer.
}
The impact of the visibility is obvious in early-fusion systems but not intermediate-fusion systems. The difference is reasonable because for early-fusion schemes, more LiDAR rays interact with closer targets thus attackers can manipulate more LiDAR points without violating LiDAR sensor physics. 

\begin{figure*}[!htb]
\minipage{0.31\textwidth}
    \includegraphics[width=\textwidth]{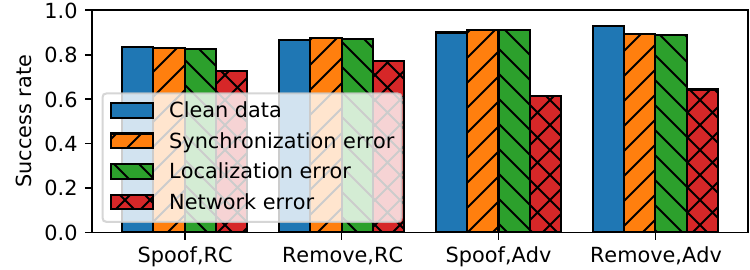}
    \vspace{-0.25in}
    \caption{\m{Attack success w/ and w/o benign errors.}}
    \vspace{-0.12in}
    \label{fig:factor_benign_error}
\endminipage\hfill
\minipage{0.32\textwidth}
    \includegraphics[width=\textwidth]{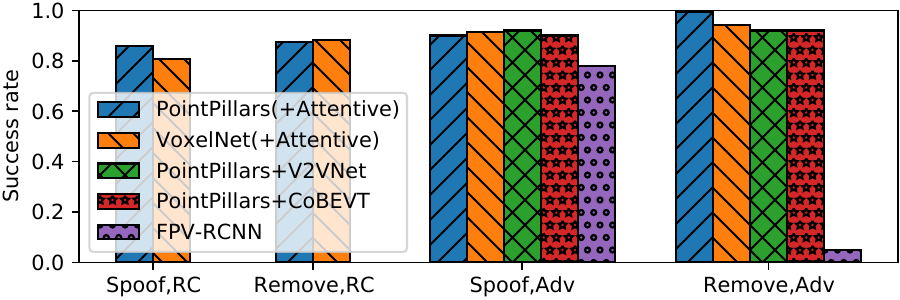}
    \vspace{-0.25in}
    \caption{\m{Attack success w.r.t. perception model configurations.}}
    \vspace{-0.12in}
    \label{fig:factor_model}
\endminipage\hfill
\minipage{0.33\textwidth}%
    \includegraphics[width=\textwidth]{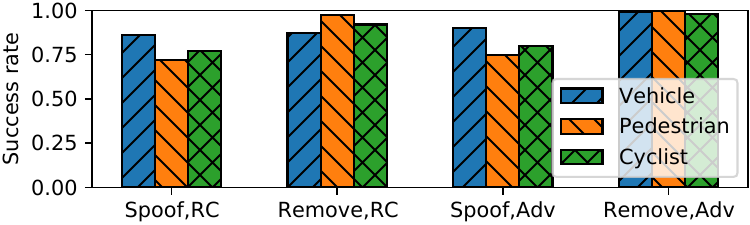}
    \vspace{-0.25in}
    \caption{\m{Attack success w.r.t. types of target objects.}}
    \vspace{-0.12in}
    \label{fig:factor_object_type}
\endminipage
\end{figure*}

\noindent
\textbf{Benign errors}. \m{It is worth nothing that attacks should tolerate errors in real systems. To simulate the worst-case synchronization errors, we delay any LiDAR frame by 100ms at a probability of 0.5.
To simulate localization errors, we incorporate uniform noise into vehicle locations (0 -- 0.2~m) and orientations (0 -- $0.2^{\circ}$), following existing works~\cite{oxts,xu2022v2x}.
Network errors can manifest as delays, corruptions, or dropped messages, all of which hinder the proper sharing of data. If the attacker's malicious data fails to reach the victim, the attack on that frame will certainly fail. Conversely, if benign vehicles' data cannot reach the attacker, less data is used for attack optimization, potentially leading to less successful attacks. To simulate such scenarios, we randomly drop 10\% of data sharing during the attacks. The 10\% error rate is regarded as the highest threshold of acceptable network connection by previous studies~\cite{toghi2018multiple,zhao2022collaborative}.
}

\m{
From the results in Figure~\ref{fig:factor_benign_error}, synchronization and localization errors have very minor impact on the attacks. Network errors decrease the success rate by 10-20\%, with a 10\% reduction attributable to the fact that 10\% of the attacker's messages fail to reach the victim. Even with the barely acceptable network connection, our attacks can achieve at least 60\% success rate, showing the robustness against benign errors.
}

\noindent
\textbf{Model configuration}. Our attacks are general for various collaborative perception models. In Figure~\ref{fig:factor_model}, the attack success rate is stable if (1) replacing the backbone model by VoxelNet~\cite{zhou2018voxelnet} in either early-fusion or intermediate-fusion methods; (2) changing the fusion network of intermediate-fusion system to V2VNet~\cite{wang2020v2vnet} \m{or CoBEVT~\cite{xu2022cobevt}. However, FPV-RCNN~\cite{yuan2022keypoints} involves a second-stage non-differential fusion on bounding box proposals (similar to late-fusion), making object removal hard.}

\noindent
\textbf{Object types}. \m{We generalize our attacks from vehicle targets to pedestrians and cyclists. As \tool{OPV2V}~\cite{xu2022opv2v} only has vehicles originally, we augment \tool{OPV2V} to include pedestrians and cyclists by modifying the simulation settings and re-training the models. As shown in Figure~\ref{fig:factor_object_type}, the attacks are generally effective for different object types. Especially, removing pedestrians is easier than removing other object types because they usually comprise a small number of LiDAR points and have a low detection confidence.}

\noindent
\textbf{Number of attackers}. One attacker is strong enough to break collaborative perception. Adding another attacker can further increase the success rate of ray casting attacks and adversarial attacks by around 5\% and 2\%, respectively.

\mysubsubsection{Ablation Study}
\label{sec:attack_ablation}



\begin{figure*}[!htb]
\minipage{0.30\textwidth}
    \includegraphics[width=\textwidth]{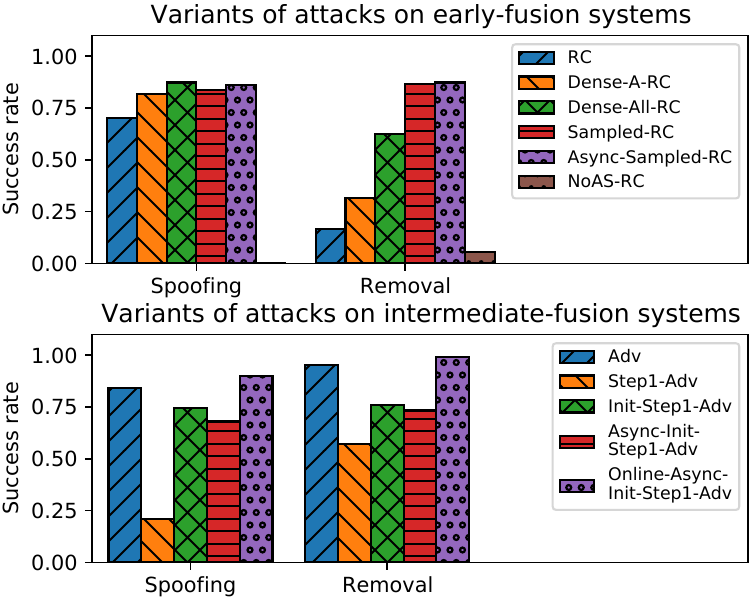}
    \vspace{-0.25in}
    \caption{Ablation study of attacks.}
    \vspace{-0.15in}
    \label{fig:ablation}
\endminipage\hfill
\minipage{0.68\textwidth}
    \includegraphics[width=\textwidth]{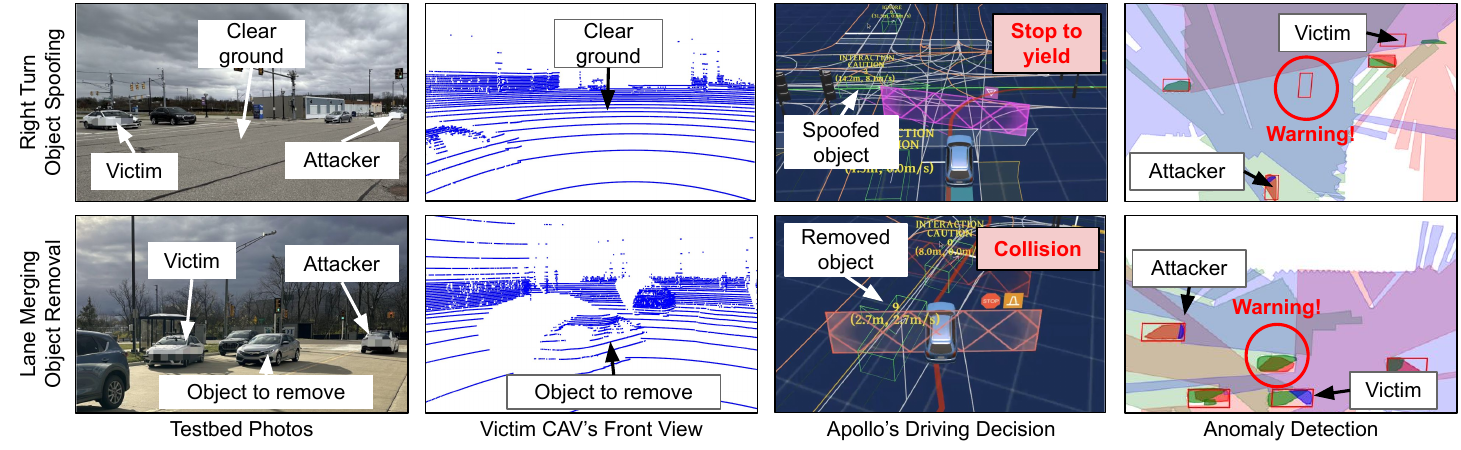}
    \vspace{-0.25in}
    \caption{Real-world experiments of attacks and anomaly detection, involving early/intermediate-fusion attacks on two scenarios.}
    \vspace{-0.15in}
    \label{fig:testbed}
\endminipage
\end{figure*}

For each attack we propose, we provide a set of variants by removing one or more components from the original design. Attack results are summarized in Figure~\ref{fig:ablation} and the complete quantitative results are in Appendix~\ref{sec:attacks_all}.

For ray casting attacks against early-fusion systems, we design the following variants.
(1) \emph{RC}. Baseline ray casting pretending the object to spoof/remove emerged/disappeared. Especially for object removal, \emph{RC} uses the adversarial shape while \emph{NoAS-RC} does not.
(2) \emph{Dense-A-RC}. Based on \emph{Naive-RC}, make the spoofed points denser by placing the origin of rays only 5-meter away from the target during ray casting.
(3) \emph{Dense-All-RC}. More than \emph{Dense-A-RC}, add multiple virtual LiDARs around the target to further increase point coverage.
(4) \emph{Sampled-RC}. Based on \emph{RC}, do non-occlusion ray casting and point sampling as mentioned in \S\ref{sec:early_fusion_attack}.
(5) \emph{Async-Sampled-RC}. The proposed attack. Based on \emph{Sampled-RC}, optimization is done one frame before the attack happens.

The attack results validate our assumptions and prove our design components are useful. 
\emph{Point density} and \emph{Point coverage} lead to stronger attacks. \emph{Dense-A-RC}'s success rate is 12\%/15\% higher than \emph{RC} for spoofing/removal. \emph{Dense-All-RC}'s success rate is 6\%/31\% higher than \emph{Dense-A-RC} for spoofing/removal.
However, \emph{Dense-A-RC} and \emph{Dense-All-RC} are not stealthy attacks as spoofed points have abnormal density.
Therefore, we propose \emph{Sampled-RC}, whose success rate is 14\%/50\% higher than the naive ray casting while preserving LiDAR's physical laws.
Finally, our asynchronous attack scheduling makes \emph{Async-Sampled-RC} deployable in real-time systems, without a significant drop in success rate.
In addition, the universal adversarial shape is crucial for object removal. Naively replacing object points with ground points only achieves a 5\% success rate and the usage of adversarial shapes raises the number to 17\%.

For adversarial attacks against intermediate-fusion systems, we design the following variants.
(1) \emph{Adv}. Basic implementation of PGD. It does not constrain the attacker's knowledge or number of optimization steps and disables black-box initialization. Instead of using perturbation masking, we add a regularization term to achieve a targeted attack.
(2) \emph{Step1-Adv}. Based on \emph{Adv}, do optimization for only one iteration. Parameters are set the same as \S\ref{sec:implementation}.
(3) \emph{Init-Step1-Adv}. Based on \emph{Step1-Adv}, add black-box initialization.
(4) \emph{Async-Init-Step1-Adv}. Based on \emph{Init-Step1-Adv}, optimization is done one frame before the attack happens.
(5) \emph{Online-Async-Init-Step1-Adv}. Our proposed attack (\S\ref{sec:intermediate_fusion_attack}). Online attack optimizing one perturbation vector over consecutive frames.

We conclude the effectiveness of our key designs.
\emph{Adv} is a standard white-box adversarial attack with the minimum constraints and maximum resources, representing the empirical upper bound of attack impact.
Limiting optimization iteration to only one per frame, though significantly lower computation cost, drops attack success rate by 63\%/48\% for spoofing/removal.
To address the problem, we propose black-box initialization. The design is very useful, especially for object spoofing: \emph{Init-Step1-Adv} achieves 53\%/8\% higher attack success rate than \emph{Step1-Adv}.
Finally, \emph{Async-Init-Step1-Adv} integrates the zero-delay attack scheduling without dropping attack effectiveness and \emph{Online-Async-Init-Step1-Adv} builds an online attack pipeline which further enhances the attacks.

\mysubsubsection{Overhead}
\label{sec:attack_performance}


We measure the execution latency of our attack algorithms in the in-vehicle execution environment.
For ray casting attacks, 3D object model preparation is done offline. The non-occlusion ray casting takes 54~ms on average. 
Our implementation of ray casting is CPU-only and can be further improved by hardware acceleration.
The point sampling takes only <3~ms. Attack transformation introduces a negligible overhead of <1~ms.
For adversarial attacks, the point cluster for black-box initialization is prepared offline thus the initialization simply appends pre-computed points to the LiDAR image, incurring a negligible overhead of <1~ms. The one-step PGD optimization is computationally intensive and requires GPU resources, taking 67~ms on average. The total attack generation is finished in 89~ms on average within one LiDAR cycle. The cost of attack transformation is negligiable.


\mysubsubsection{Real-world Case Study}
\label{sec:attack_testbed}


Attacks must be realizable. We test attack algorithms by emulating driving scenarios using dataset \tool{Adv-\testbed}. In this section, we focus on case studies on two scenarios, as shown in Figure~\ref{fig:testbed}. All scenarios are described in Appendix~\ref{sec:testbed}.

\emph{Object spoofing during right turn}. The victim CAV is turning right at green while the attacker CAV stops on another road. The attacker's goal is to spoof one fake vehicle to stop the victim, forming a denial-of-service (DoS). First, we launch ray casting attack assuming CAVs use early-fusion collaborative perception. Since the victim is far away from the attacker (>30~m), it is hard to directly spoof an object in front of the victim. However, the attacker can leverage the traffic rule implemented in CAVs by spoofing a moving vehicle whose trajectory blocks the victim's path. In 5 seconds, the attack succeeds to spoof the vehicle in 76\% of frames.
Baidu Apollo indeed stops the vehicle to yield the spoofed vehicle. 
Second, if CAVs use an intermediate-fusion system, the adversarial attack can achieve a stronger attack by spoofing an obstacle right in front of the victim in 92\% of frames.

\emph{Object removal during lane merging}. The victim CAV is starting from a parking place to merge into the main road while another vehicle is going through from behind. Normally, the victim should yield the right of way. The attacker sits on another lane, aiming to remove the moving vehicle from the view of the victim. The ray casting attack succeeds in removing the vehicle in the first 45 frames but fails in the last 5 frames because the target is further. Nevertheless, it is too late when the victim perceives the target and Baidu Apollo reports a collision. Also, using the white-box adversarial attack against intermediate-fusion perception has a similar attack impact, removing the vehicle in 96\% of frames.


\mysubsection{Evaluation of Anomaly Detection}
\label{sec:defense_eval}


We evaluate effectiveness and efficiency in \S\ref{sec:defense_results} and \S\ref{sec:defense_overhead}. We then compare \system with existing defenses in \S\ref{sec:defense_compare}. We demonstrate the real-world deployment in \S\ref{sec:defense_testbed}.

\mysubsubsection{Defense Results}
\label{sec:defense_results}


\begin{figure}[t]
   \begin{minipage}{0.23\textwidth}
     \centering
     \includegraphics[width=.9\linewidth,height=.8\linewidth]{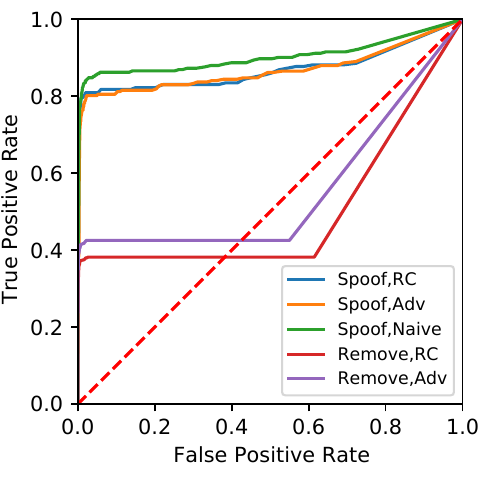}
     \vspace{-0.06in}
     \caption{ROC curve of anomaly detection.}
     \label{fig:defense_results_roc}
     \vspace{-0.05in}
   \end{minipage}\hfill
   \begin{minipage}{0.23\textwidth}
     \centering
     \includegraphics[width=.9\linewidth,height=.8\linewidth]{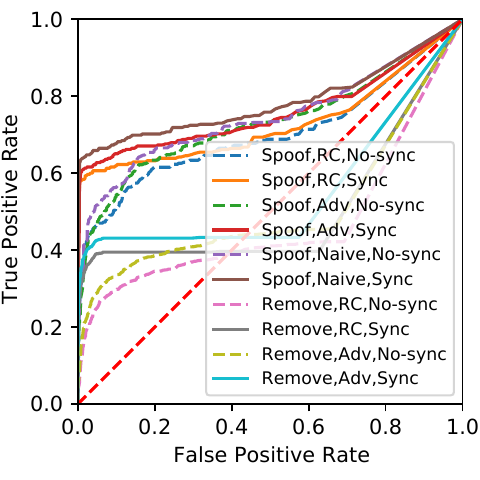}
     \vspace{-0.06in}
     \caption{ROC curve w/ and w/o synchronization.}
     \label{fig:defense_sync}
     \vspace{-0.05in}
   \end{minipage}
\end{figure}



\begin{figure*}[!htb]
\minipage{0.32\textwidth}
    \includegraphics[width=\textwidth]{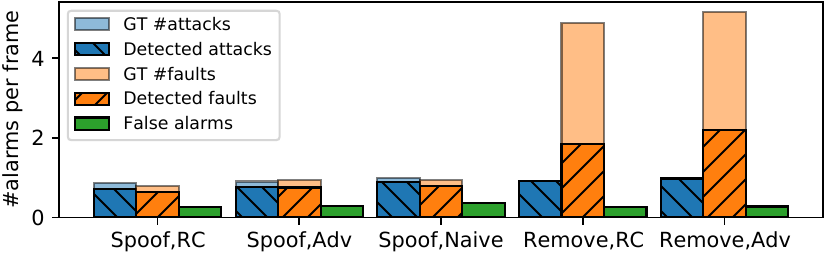}
    \vspace{-0.25in}
    \caption{Split-down of alarms of anomaly detection.}
    \label{fig:defense_results_bar}
    \vspace{-0.15in}
\endminipage\hfill
\minipage{0.32\textwidth}
    \includegraphics[width=\textwidth]{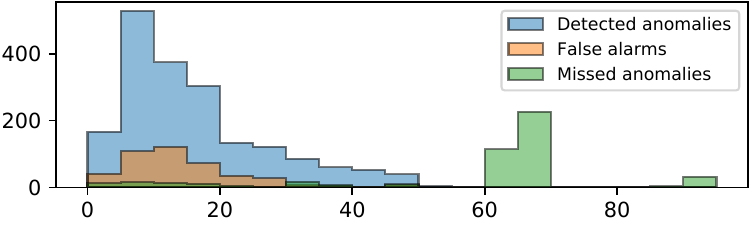}
    \vspace{-0.25in}
    \caption{\#Alarms w.r.t. distance to benign LiDARs.}
    \label{fig:defense_factor_distance}
    \vspace{-0.15in}
\endminipage\hfill
\minipage{0.33\textwidth}%
    \includegraphics[width=\textwidth]{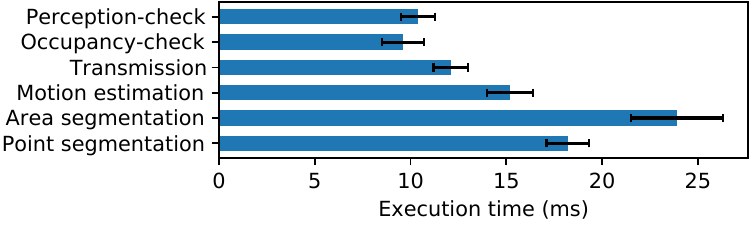}
    \vspace{-0.25in}
    \caption{Latency of anomaly detection.}
    \label{fig:defense_overhead}
    \vspace{-0.15in}
\endminipage
\end{figure*}

We apply \system on attacked frames in \tool{Adv-OPV2V}. Note that \system is supposed to detect both attacks and perception faults, as long as the predicted bounding box has no overlap with ground-truth or the ground-truth bounding box is not detected. We consider adaptive attacks, where the attacker fakes his/her occupancy map to avoid conflicts with other occupancy maps or detected bounding boxes. Therefore, \emph{occupancy consistency check} in \S\ref{sec:defense_design} cannot defend adaptive attacks but serves as input validation before merging occupancy maps.

\m{
In Table~\ref{tab:attack_results}, true positive rate (TPR) and false positive rate (FPR) are calculated on the whole LiDAR images, including the detection of both malicious attacks and benign perception faults.
On the other hand, success rate measures the detection of only malicious attacks, which is the ratio of positive detection on the target region and the total number of attack scenarios.
We also show the ROC curves in Figure~\ref{fig:defense_results_roc}.
}
\system is generally effective against various attack methods. If selecting thresholds $\sigma_{spoof}$ and $\sigma_{remove}$ to maximize AUC score, \system achieves FPR <3\% and TPR >80\%/38\% against spoofing/removal while detecting around 90\% of anomalies caused by our attacks.
From the split-down of alarms in Figure~\ref{fig:defense_results_bar}, low TPR against removal threats is mainly caused by undetected benign perception faults which are out of the range of occupancy maps.
The ``false alarms'' are mostly the cases where predicted bounding boxes is not accurate (IoU < 0.5). Though considered as normal cases by our criteria, they have significant differences with accurate object detection.

\m{We also apply \system on the prior attack~\cite{tu2021adversarial}. As the prior attack is untargeted and injects a few dozens of fake detection results in each LiDAR image, it is easy for \system to reveal 100\% of the attacked frames, as shown in Table~\ref{tab:attack_results}. The low TPR (around 30\%) is because the occupancy maps cannot cover lots of the far-away fake detection results.}

\noindent
\textbf{Other adaptive attacks}. The attacker may exploit \emph{occupancy consistency check} to create as many conflicts as possible to minimize the coverage of the merged occupancy map and decrease TPR. However, if the occupancy conflict is with the victim's local map, the attacker is directly identified because the local data is trusted. This ensures a lower bound of TPR by using only local occupancy maps (71.7\%/24.0\%/70.4\%/28.7\%/78.1\% against the attacks in Table~\ref{tab:attack_results}). Also, occupancy conflicts obviously indicate the existence of attackers and are useful messages for other defense mechanisms such as reputation systems.

The attacker may also choose to launch attacks at locations out of the coverage of occupancy maps. However, our experiments on \tool{Adv-OPV2V} show that benign occupancy maps cover 95.6\% in 30 meters and 99.9\% in 10 meters around the victim. It is very little chance for the attacker to spoof/remove objects stealthily at a safety-critical distance.

\mysubsubsection{Impacting Factors}
\label{sec:defense_factors}


\noindent
\textbf{Distance to LiDAR sensors}. As shown in Figure~\ref{fig:defense_factor_distance}, over 80\% false alarms are 60 meters away from any benign vehicles. Within the range of occupancy maps (50 meters in our configuration), \system can stably make true detection.

\noindent
\textbf{Synchronization}. With injected synchronization errors as introduced in \S\ref{sec:attack_factors}, \system's synchronization provides significant robustness. 
As shown in Figure~\ref{fig:defense_sync}, \system is not effective without synchronization, having TPR 35\%/15\% against spoofing/removal when FPR is low (<5\%). With synchronization, \system achieves TPR 60\%/40\% against spoofing/removal, close to the detection rate on ideal synchronized data.

\noindent
\textbf{Localization errors}. With the injected localization errors as stated in \S\ref{sec:attack_factors}, we observe a minor decrease of accuracy (TPR -3.1\%, FPR +0.2\%), showing \system's robustness.

\noindent
\textbf{Object types}. \m{As \system uses the area of conflicted regions as the key metric, smaller object sizes result in higher FPR, e.g., minor conflicts caused by errors of occupancy maps may be falsely considered as anomalies. By choosing the best AUC score of the ROC curve, \system detects pedestrian spoofing, pedestrian removal, cyclist spoofing, and cyclist removal in FPR/TPR of 78.4\%/14.5\%, 38.2\%/13.9\%, 81.7\%/6.5\%, and 29.4\%/6.2\%, respectively. When compared with the detection of fake vehicles, \system yields around 12\%/4\% higher FPR on pedestrians/cyclists while maintains TPR stable.}

\noindent
\textbf{Number of attackers}. 
More attackers decrease the coverage of benign occupancy maps and cause more false negatives. Adding another attack decreases TPR by 5\% on average.

\mysubsubsection{Overhead}
\label{sec:defense_overhead}


We measure the latency of the anomaly detection using the in-vehicle execution environment and recorded network traces~\cite{narayanan2021variegated}. Segmentation/clustering algorithms are relatively expensive but they can be further boosted using hardware acceleration~\cite{canilho2016multi}. Occupancy map transmission is as fast as 10ms. Each map contains around 300-1000 polygon vertices and lightweight metadata of object motion, in a small size of around 10 KB. 
Consistency checks are simple polygon operations that can be finished in 15ms.
The end-to-end anomaly detection takes 92ms, which means the CAV can be aware of abnormal bounding boxes before the LiDAR cycle ends.

\mysubsubsection{Comparison with Other Defense Approaches}
\label{sec:defense_compare}

\begin{figure}[t]
   \begin{minipage}{0.15\textwidth}
     \centering
     \includegraphics[width=.995\linewidth]{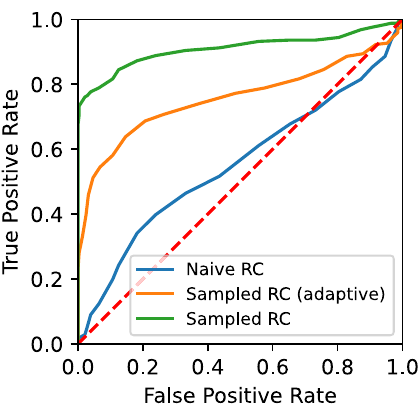}
     \vspace{-0.20in}
     \caption{ROC curve of \tool{CARLO}.}
     \label{fig:compare_carlo}
     \vspace{-0.05in}
   \end{minipage}
   \begin{minipage}{0.15\textwidth}
     \centering
     \includegraphics[width=.995\linewidth]{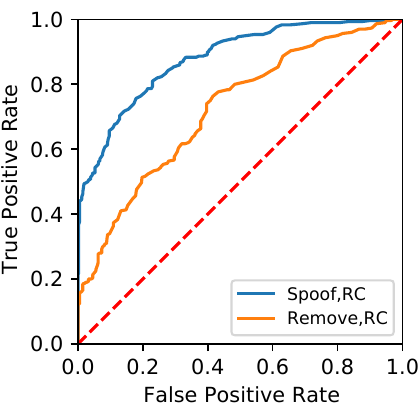}
     \vspace{-0.20in}
     \caption{ROC curve of \tool{LIFE}.}
     \label{fig:compare_life}
     \vspace{-0.05in}
   \end{minipage}
   \begin{minipage}{0.15\textwidth}
     \centering
     \includegraphics[width=.995\linewidth]{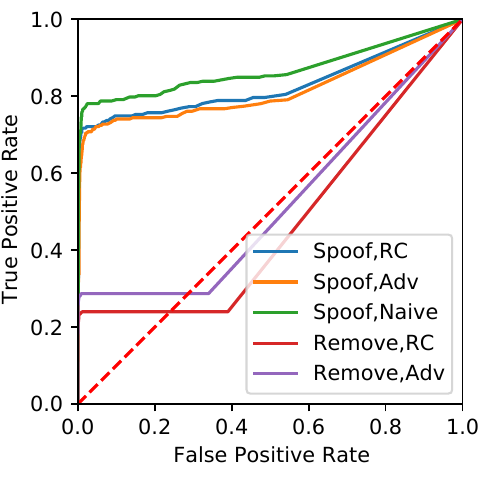}
     \vspace{-0.20in}
     \caption{\m{ROC curve of \tool{MDS}.}}
     \label{fig:compare_mds}
     \vspace{-0.05in}
   \end{minipage}
\end{figure}


\noindent
\textbf{CARLO}~\cite{sun2020towards} is an anomaly detection algorithm detecting LiDAR spoofing attacks. Given the fact that detected bounding boxes should host solid objects, \tool{CARLO} validates that the volume of ``free space'' (conical spaces between the LiDAR sensor and rendered points) in bounding boxes is under a threshold. In collaborative perception, we assume the victim CAV applies \tool{CARLO} on each received LiDAR point cloud. 

Results are shown in Figure~\ref{fig:compare_carlo}, which detects our proposed ray casting attack (Sampled RC) with TPR 77.7\% and FPR 3.9\%. However, attackers can adjust the rule of point sampling in \S\ref{sec:early_fusion_attack} to launch adaptive attacks. For instance, the attacker can restrict the number of points that can penetrate object surfaces to be <30\% (Sampled RC adaptive). As a result, TPR decreases to 63.8\% and FPR increases to 14.7\% while the success rate only drops by 5.6\%. If forbid ray penetration completely (Naive RC), \tool{CARLO} is close to random guessing while the success rate of our attack is still above 70\%. In contrast, \system achieves higher TPR and, more importantly, is independent of attack methods.


\noindent
\textbf{LIFE}~\cite{liu2021seeing} is a hybrid anomaly detection system against sensor attacks. 
First, it checks the temporal consistency of depth camera images based on machine learning methods. As discussed in \S\ref{sec:defense_challenge}, attackers have the capability to continuously launch attacks thus the check is fundamentally not useful.
Besides, an object matching algorithm checks the consistency between objects detected in the camera and the LiDAR. In early-fusion systems, CAVs can launch object matching on remote LiDAR and local camera images. To reproduce \tool{LIFE}, we use the same LiDAR segmentation as \system and train a \tool{EfficientPS}~\cite{mohan2021efficientps} model for camera image segmentation.

We draw the ROC curve of object matching in Figure~\ref{fig:compare_life}. \tool{LIFE}'s object matching achieves around 80\% TPR and 26\% FPR against early-fusion ray casting attacks. \tool{LIFE} suffers from a higher FPR because multiple machine learning processes introduce more errors - inaccurate detection from either camera or LiDAR, which is usual on far-away objects, may trigger a false alarm. 
Compared with \tool{LIFE}, \system has a higher detection rate with much lower computation/bandwidth consumption, thanks to the collaboration among CAVs.

\noindent
\textbf{MDS}~\cite{ambrosin2019design} is an anomaly detection framework assuming CAVs to share bounding boxes. Besides checks on message format and temporal consistency which are not relevant to our attacks, each CAV evaluates the consistency between the local occupancy map and final perception results, and also merges anomaly detection results from multiple CAVs by majority voting. However, the attackers can launch adaptive attacks to only send falsified data to the specific victim instead of all other CAVs, thus the majority voting is actually not helpful.
Compared with \system, the spatial check is restricted on the local occupancy map (without the sharing of occupancy maps) thus TPR is lower by 9-15\%, as shown in Figure~\ref{fig:compare_mds}.

\system has no conflicts with the above defenses. Users can deploy multiple defenses to strengthen sensor data integrity.

\mysubsubsection{Real-world Case Study}
\label{sec:defense_testbed}

We demonstrate \system on the same attack scenarios discussed in \S\ref{sec:attack_testbed}, shown in Figure~\ref{fig:testbed}.
In the scenario of the right turn, though the spoofed object is in the blind spot of the victim (red), another benign vehicle (blue) observes that region and identifies the anomaly when it is 15 meters away from the victim.
In the scenario of lane merging, the victim vehicle recognizes an object point cluster on the left but is not detected by the perception system, triggering a warning of object removal 2.1 seconds before a potential collision.
In other scenarios in Appendix~\ref{sec:testbed}, our anomaly detection can detect attacks at least 1.5 seconds before a collision or hard brake happens. 
The anomaly detection can be more robust when there are more benign vehicles on busy roads. 


\mysection{Conclusion}

In this work, we pioneer to examine the threats posed by data fabrication on collaborative perception systems. We unleash novel attacks that successfully spoof or remove on-road objects in various types of collaborative perception schemes and demonstrate the attack impact on real traffic scenarios. To mitigate the threats, we introduce a cross-vehicle validation solution powered by fine-grained occupancy maps, which detects anomalies seconds before potential road hazards occur. Our attempts of both attacks and defenses serve as a benchmark to spur future research on collaborative perception security.

\vspace{-1em}
\section*{Acknowledgments}
\vspace{-0.4em}
We would like to thank our anonymous shepherd and reviewers for their valuable comments and feedback. This work was supported in part by NSF under CNS-1930041, CMMI-2038215, CNS-1932464, CNS-1929771, and CNS-2145493, USDOT CARMEN University Transportation Center (UTC), and the National AI Institute for Edge Computing Leveraging Next Generation Wireless Networks, Grant \# 2112562.


\bibliographystyle{plain}
{\footnotesize
\bibliography{reference}}

\appendix
\mysection{Algorithm Details}

We introduce more implementation details of attack and defense algorithms proposed in \S\ref{sec:attack_design} and \S\ref{sec:defense_design}.

\mysubsection{Universal Adversarial Shape}
\label{sec:adv_shape}

As mentioned in \S\ref{sec:early_fusion_attack}, the universal adversarial shape is for early-fusion object removal attacks. We find that spoofing an adversarial shape covering the original object to remove is surprisingly useful. As the generation of adversarial shapes involves time-consuming optimization, we aim to generate a universal shape that can be pre-computed offline and is general for different scenarios.

\begin{figure}[t]
    \centering
    \includegraphics[width=0.2\textwidth]{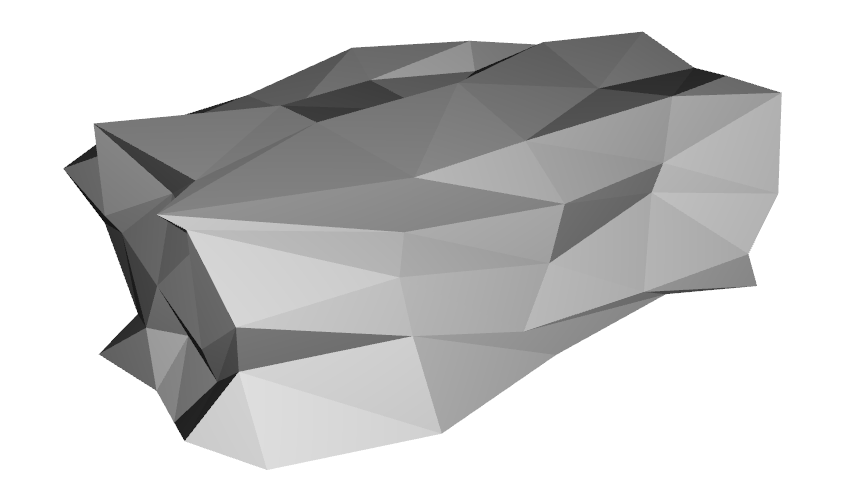}
    \vspace{-0.10in}
    \caption{Example of universal adversarial shape.}
    \vspace{-0.01in}
    \label{fig:adv_shape}
\end{figure}

As shown in Algorithm~\ref{alg:adv_shape_appendix}, the generation algorithm is based on black-box genetic optimization. In our implementation, we create the initial shape as a cuboid and randomly select 50 diverse object removal attack scenarios as $X$. For the genetic algorithm, the initial population is a set of random perturbations on the initial shape and other parameters are set as mentioned in \S\ref{sec:implementation}. For each perturbed shape, we launch the attack defined in \S\ref{sec:early_fusion_attack} on each attack case and calculate an average fitness score as Equation~\ref{eq:adv_shape}: 

\vspace{-1em}
\begin{equation}
\label{eq:adv_shape}
F(y) = \sum_{y\in Y} \textproc{IoU}(y, y_t) \cdot log(1 - y_\sigma),
\end{equation}
\vspace{-1em}

where $y\in Y$ denotes predicted bounding boxes, $y_\sigma$ denotes confidence scores, and $y_t$ denotes the target region. The black-box optimization minimizes Intersection over Union (IoU) on the target region and confidence score simultaneously.
Note that we will scale and transform the shape to a bounding box that has the same center as the target region but is 0.6~m (fine-tuned by experiments) larger on all dimensions, making sure the adversarial shape covers the original object edges. The remaining optimization steps are handled by the genetic algorithm itself. Figure~\ref{fig:adv_shape} shows the universal adversarial shape we used for attacking \tool{Adv-OPV2V}.

\begin{algorithm}[t]
\tiny
\SetAlgoLined
\KwIn{An initial 3D shape $S_0$, a set of attack cases $X$.}
\KwOut{A universal adversarial shape $S_{adv}$.}
Initialize a set of perturbation on vertices of $S_{in}$: $P$\;
Initialize genetic algorithm instance: $\textproc{Genetic}(P)$\;
\For {Iteration 1\dots K} {
    \For {$p_i \in P$} {
        \For {$x_j \in X$} {
            $x_j' \gets \textproc{RemovalAttack}(x_j, S_0 + p_i)$\;
            $Y_i' \gets \textproc{Perception}(x_j')$\;
            $I_i \gets I_i - \sum_{y\in Y_i'} \textproc{IoU}(y, y_t^{(j)}) \cdot log(y_\sigma)$\;
        }
    }
    $S_{adv} \gets S_0 + \textproc{BestSolution}(P, I)$
    $P \gets \textproc{GeneticUpdate}(P, I)$\;
}
 \caption{Universal adversarial shape generation.}
 \label{alg:adv_shape_appendix}
\end{algorithm}

\mysubsection{Point Sampling}
\label{sec:point_sampling}

The point sampling resolves occlusion conflicts after ray casting and meanwhile maximizes attack effectiveness, as shown in Algorithm~\ref{alg:point_sampling}. Each ray after non-occlusion ray casting has a set of intersection points with the target 3D model and the point sampling algorithm assigns each intersection point with a probability, which is larger when the points are closer to benign LiDARs, as mentioned in \S\ref{sec:early_fusion_attack}. 
The location of LiDARs is broadcast in collaborative perception, which is necessary to merge multi-source LiDAR images.
Also, there is a parameter tuning how much rays can penetrate object surfaces. The fewer rays that can penetrate surfaces, the more natural the spoofed point clouds are, as discussed in \S\ref{sec:defense_compare}.

\begin{algorithm}[t]
\tiny
\SetAlgoLined
\KwIn{A mapping from one ray to a set of points intersected with 3D model $M$, LiDAR poses of benign vehicles $L$, a parameter of the probability of penetration $0 \le \sigma_{p} \le 1$.}
\KwOut{A mapping from one ray to one intersection point $M'$.}
\For{$r, X \in M$} {
    \eIf{$\textproc{Random}() < \sigma_{p}$} {
        $D \gets \textproc{ComputeDistance}(X, L)$\;
        $P \gets \textproc{Probability}(D)$\;
        $M'(r) \gets \textproc{RandomChoice}(X, P)$\;
    } {
        $M'(r) \gets \textproc{Closest}(X)$\;
    }
}
 \caption{Point Sampling.}
 \label{alg:point_sampling}
\end{algorithm}






\mysubsection{Space Segmentation}
\label{sec:space_seg}

We provide implementation details for identifying free regions to the best precision.

\begin{figure}[t]
    \centering
    \includegraphics[width=0.4\textwidth]{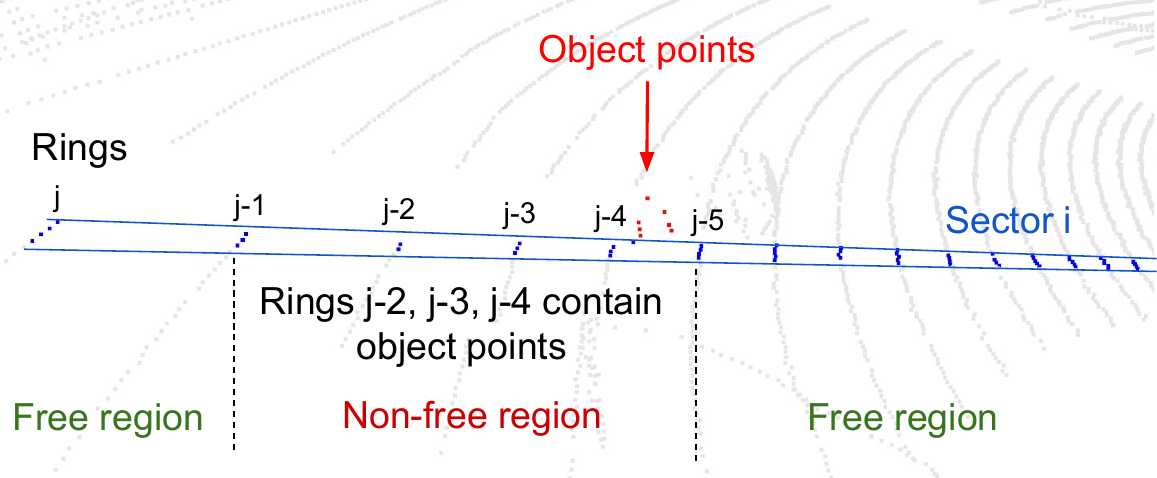}
    \vspace{-0.15in}
    \caption{Example of free space identification.}
    \vspace{-0.02in}
    \label{fig:space_seg}
\end{figure}

\begin{algorithm}[t]
\tiny
\SetAlgoLined
\KwIn{Point cloud $X$, rings $R$, sectors $E$, object points $X_{obj} \subseteq X$, ground points $X_{grd} \subseteq X$.}
\KwOut{Polygons $S_F$ as 2D free regions.}
$S_F \gets \emptyset$\;
$F \gets \emptyset$ as the set of ground point groups\;
\For {$e_i \in E$} {
    \For {$r_j \in R$} {
        $X_{i,j} \gets \textproc{GetPoints}(e_i, r_j)$\;
        \If {$X_{i,j} \cap X_{obj} = \emptyset$ and $X_{i,j} \subseteq X_{grd}$} {
            Push $X_{i,j}$ to $F$\;
        } 
    }
    \While {Find $r_{j}, r_{j+1}, \dots, r_{j+k} \in F$} {
        Get $P$ as a polygon in sector $e_i$ and rings $r_{j},r_{j+k}$\;
        \If {Cannot ind $x \in X_{obj}$ in $P$} {
            Push $P$ to $S_F$\;
        }
    }
}
 \caption{Segmentation of free regions.}
 \label{alg:space_seg}
\end{algorithm}

LiDAR sensors equip vertically placed laser transmitters. In each LiDAR cycle, the LiDAR sensor rotates itself thus the lasers scan the object surfaces. Therefore, points in the LiDAR image can be grouped according to which laser they belong to, called ``rings''. Inner rings are closer to the LiDAR sensor, and vice versa. Also, we can group LiDAR points by splitting the bird-eye view 2D space into sectors whose vertex is the location of the LiDAR sensor. In this way, each LiDAR point has a ring ID and a sector ID, as shown in Figure~\ref{fig:space_seg}.
After grouping LiDAR points into rings and sectors, we execute Algorithm~\ref{alg:space_seg} to label free regions. First of all, we identify point groups that entirely consist of ground points. Then, we search in each sector $i$ for consecutive rings from $j$ to $j+k$ which are all ground point groups and generate a polygon covering a segment of sector $j$ between the furthest point of ring $i$ and the closest point of ring $j+k$. For now, we ensure that this region is not occluded by other objects, otherwise at least one point in these rings is an object point.
Next, we double check no object point falls in the polygon, in case there are objects above the ground.
If no object points are found, we can finally label the polygon as a free region. 

Our free space segmentation is precise in two folds. (1) The grouping of LiDAR points follows the physical laws of LiDAR sensors thus each group has a similar number of points, making the segmentation stable even in far-away regions. (2) The identification of free regions is conservative by taking occlusion and floating objects into account.

Note that the size of sectors and rings is configurable for a good trade-off between precision and data size. In our implementation, we split the 2D space into 360 sectors, each for $1^\circ$. The number of rings depends on the number of lasers the LiDAR sensor has, for instance, LiDARs in \tool{Adv-OPV2V} have 64 lasers while LiDARs in \tool{Adv-\testbed} have 32 lasers.

\mysection{Testbed \testbed}
\label{sec:testbed}

\begin{figure}[t]
    \centering
    \includegraphics[width=0.48\textwidth]{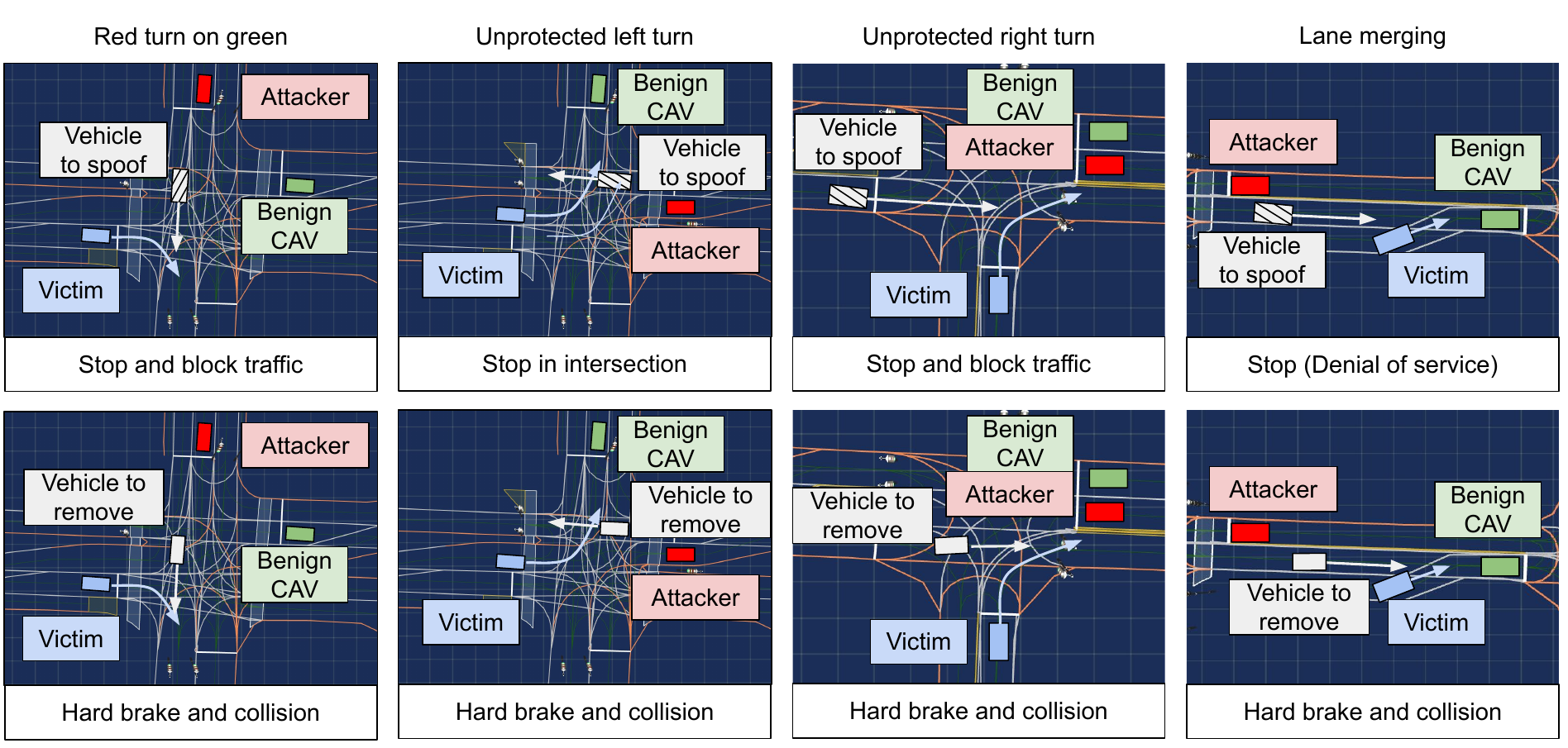}
    \vspace{-0.15in}
    \caption{8 attack scenarios in \tool{Adv-\testbed}.}
    \vspace{-0.03in}
    \label{fig:testbed_all}
\end{figure}

Figure~\ref{fig:testbed_all} depicts the 8 attack scenarios we created in Testbed \testbed. We show the topology of on-road vehicles as well as expected attack impacts.
We create 4 traffic scenes: (1) a right turn on green in an intersection, (2) an unprotected left turn in an intersection, (3) an unprotected right turn in a T intersection, and (4) a lane merging from a parking place. 
Each scene contains one attacker CAV, one victim CAV, one benign CAV, and 1-2 regular vehicles.
In each traffic scene, we craft one object spoofing attack case and one object removal attack case. Object spoofing intends to spoof a vehicle to the view of the victim. It aims to stop the victim from moving for as long time as possible, forming a denial of service. Object removal aims to remove an on-road vehicle, which is important for driving safety, from the victim's perception results. Therefore, object removal tends to trigger severe safety hazards such as collisions.
We recorded 5-second LiDAR/GPS packets and network traces for each scenario to emulate attacks.

In experiments, we launch both black-box ray casting attacks (\S\ref{sec:early_fusion_attack}) and white-box adversarial attacks (\S\ref{sec:intermediate_fusion_attack}) on 8 scenarios, assuming the CAVs may host either early-fusion or intermediate-fusion collaborative perception.
All attacks impact (e.g., stop or collision) are successfully produced.
Among the 400 frames of perception (5 seconds of 10 Hz LiDAR images for 8 scenarios), the ray casting attacks succeed in object spoofing/removal in 87\%/79\% of frames while the adversarial attacks succeed in object spoofing/removal in 92\%/95\% of frames.
Adversarial attacks in general have stronger attack effects as their success rate is not restricted by the distance between targets and the attacker.

We also launch an anomaly detection \system on attacked data, achieving 85.3\% TPR and 2.6\% FPR on average. The alarms are useful to avoid safety hazards as they are delivered 1.5 seconds on average before collisions or brakes happen.

\begin{table}[t]
\tiny
\caption{Attack variants evaluated on \tool{Adv-OPV2V}.}
  \label{tab:attacks_all}
  \centering
  \begin{tabular}{| c | c | l | c | c | c | c |}
    \noalign{\global\arrayrulewidth1pt}\hline\noalign{\global\arrayrulewidth0.4pt}
    Fusion & Goal & Method & Success & IoU & Score & $\Delta$AP \\
    \noalign{\global\arrayrulewidth1pt}\hline\noalign{\global\arrayrulewidth0.4pt}
    Early & Spoof & RC & 70.3\% & 0.39 & 0.29 & -0.39\% \\
    Early & Spoof & RC,Dense-A & 81.7\% & 0.46 & 0.35 & -0.44\% \\
    Early & Spoof & RC,Dense-All & 87.3\% & 0.54 & 0.39 & -0.48\% \\
    Early & Spoof & RC,Sampled & 83.7\% & 0.53 & 0.37 & -0.39\% \\
    Early & Spoof & RC,Sampled,Async & 86.0\% & 0.54 & 0.38 &  -0.42\% \\
    Early & Remove & RC & 5.3\% & 0.73 & 0.45 & -0.09\% \\
    Early & Remove & RC,Dense-A & 6.3\% & 0.70 & 0.44 & -0.05\% \\
    Early & Remove & RC,Dense-All & 9.7\% & 0.66 & 0.41 & -0.05\% \\
    Early & Remove & RC,AS & 6.7\% & 0.53 & 0.38 & -0.08\% \\
    Early & Remove & RC,AS,Dense-A & 21.7\% & 0.50 & 0.31 & -0.05\% \\
    Early & Remove & RC,AS,Dense-All & 62.3\% & 0.25 & 0.15 & -0.09\% \\
    Early & Remove & RC,AS,Sampled & 86.7\% & 0.07 & 0.04 & -0.48\% \\
    Early & Remove & RC,AS,Sampled,Async & 87.3\% & 0.07 & 0.03 & -0.49\% \\
    Intermediate & Spoof & RC & 36.3\% & 0.13 & 0.18 & -0.14\% \\
    Intermediate & Spoof & RC,Dense-A & 36.7\% & 0.14 & 0.18 & -0.11\% \\
    Intermediate & Spoof & RC,Dense-All & 42.7\% & 0.18 & 0.21 & -0.15\% \\
    Intermediate & Spoof & Adv & 60.3\% & 0.29 & 0.56 & -3.60\% \\
    Intermediate & Spoof & Adv,Step1 & 21.0\% & 0.02 & 0.12 & -0.84\% \\
    Intermediate & Spoof & Adv,Step1,Async & 26.3\% & 0.05 & 0.14 & -0.85\% \\
    Intermediate & Spoof & Adv,Step1,Async,Online & 56.7\% & 0.23 & 0.42 & -5.36\% \\
    Intermediate & Spoof & Adv,Step1,Init & 74.7\% & 0.43 & 0.48 & -0.73\% \\
    Intermediate & Spoof & Adv,Step1,Async,Init & 68.0\% & 0.33 & 0.41 & -0.99\% \\
    Intermediate & Spoof & Adv,Step1,Async,Online,Init & 90.0\% & 0.46 & 0.70 & -2.01\% \\
    Intermediate & Remove & RC,AS & 3.7\% & 0.81 & 0.56 & -0.12\% \\
    Intermediate & Remove & RC,AS,Dense-A & 6.0\% & 0.81 & 0.52 & -0.05\% \\
    Intermediate & Remove & RC,AS,Dense-All & 9.7\% & 0.81 & 0.51 & -0.04\% \\
    Intermediate & Remove & Adv & 78.7\% & 0.81 & 0.07 & -0.07\% \\
    Intermediate & Remove & Adv,Step1 & 57.3\% & 0.81 & 0.18 & -1.38\% \\
    Intermediate & Remove & Adv,Step1,Async & 56.0\% & 0.81 & 0.19 & -1.45\% \\
    Intermediate & Remove & Adv,Step1,Async,Online & 54.0\% & 0.81 & 0.17 & -1.38\% \\
    Intermediate & Remove & Adv,Step1,Init & 55.0\% & 0.81 & 0.17 & -1.25\% \\
    Intermediate & Remove & Adv,Step1,Async,Init & 57.7\% & 0.81 & 0.17 & -2.63\% \\
    Intermediate & Remove & Adv,Step1,Async,Online,Init & 93.0\% & 0.81 & 0.02 & -3.90\% \\
    Late & Spoof & Naive & 98.7\% & 0.96 & 0.99 & 0 \\
    Late & Remove & Naive & 0.3\% & 0.77 & 0.59 & 0 \\
    \noalign{\global\arrayrulewidth1pt}\hline\noalign{\global\arrayrulewidth0.4pt}
  \end{tabular}
\end{table}

\mysection{Results of Attack Variants}
\label{sec:attacks_all}

In Table~\ref{tab:attacks_all}, We list the quantified attack impact of all attack variants mentioned in our ablation study~\ref{sec:attack_ablation}, including evaluation metrics in \S\ref{sec:attack_results}: success rate, IoU, confidence score, and $\Delta$AP. All results in Table~\ref{tab:attacks_all} are run on \tool{OPV2V}'s pre-trained models, using \tool{PointPillars} as the backbone. Similar to \S\ref{sec:attack_ablation}, the following abbreviation words stand for various design choices:

\BULLET \emph{RC}. Ray casting algorithm to reconstruct point clouds.

\BULLET \emph{Adv}. Adversarial machine learning to optimize attacks.

\BULLET \emph{Async}. Use the data in frame $i-1$ to attack frame $i$.

\BULLET \emph{AS}. Universal adversarial shape generated offline, used in ray casting.

\BULLET \emph{Dense-A}. During ray casting, increase the point density of the point cloud in the target region, by moving the origin of LiDAR rays closer to the target.

\BULLET \emph{Dense-All}. During ray casting, increase the point density of the point cloud by pretending to have multiple LiDARs from different angles.

\BULLET \emph{Sampled}. First perform non-occlusion ray casting  and then sample points to resolve occlusion conflicts.

\BULLET \emph{Step1}. During adversarial attacks, do one-step PGD instead of unlimited iterations in each LiDAR cycle.

\BULLET \emph{Init}. During adversarial attacks, initialize the point cloud using a dense point cluster generated offline.

\end{document}